\documentclass[12pt]{iopart}
\usepackage{iopams}  
\usepackage{hyperref}  
\usepackage{multicol}  
\usepackage{graphicx}
\usepackage{xcolor}

\newcommand{\pd}[2]{\frac{\partial #1}{\partial #2}}
\newcommand{\gyav}[1]{\left\langle #1 \right\rangle_\mathbf{R}}
\newcommand{\gyavalt}[1]{\left\langle #1 \right\rangle_\mathbf{r}}
\newcommand{\fsav}[1]{\left\langle #1 \right\rangle_\psi}
\newcommand{\vd}{\mathbf{v_D}}
\newcommand{\vchi}{\mathbf{v}_\chi}
\newcommand{\intd}[1]{ \mathrm{d} #1 \,}
\newcommand{\dv}{\mathrm{d}^3\mathbf{v} }
\newcommand{\taueff}{\tau_\mathrm{eff}}
\newcommand{\betaeff}{\beta_\mathrm{eff}}
\newcommand{\corr}[1]{ {#1}}

\newcommand{\ian}[1]{ {\color{purple}\textbf{ #1}}}

\begin{document}
\title[Effect of fast ions on microturbulence]{First principles of modelling the stabilization of microturbulence by fast ions}
%\title[Fast ion stabilization of ITG]{Theoretical basis for the effect of fast ions on microturbulence}

\author{G. J. Wilkie, A. Iantchenko, I. G. Abel, E. Highcock,  I. Pusztai, JET Contributors\cite{litaudon_overview_2017}}

\address{Department of Physics, Chalmers University of Technology,
  Gothenburg, SE-41296, Sweden}

\begin{abstract}
   The observation that fast ions stabilize ion-temperature-gradient-driven microturbulence has profound implications for future fusion reactors. It is also important in optimizing the performance of present-day devices. 
	In this work, we examine in detail the phenomenology of fast ion stabilization and present a reduced model which describes this effect.
	This model is derived from the high-energy limit of the gyrokinetic equation and extends the existing ``dilution'' model to account for nontrivial fast ion kinetics.
	Our model provides a physically-transparent explanation for the observed stabilization and makes several key qualitative predictions.
	Firstly, that different classes of fast ions, depending on their radial density or temperature variation, have different stabilizing properties. 
	Secondly, that zonal flows are an important ingredient in this effect precisely because the fast ion zonal response is negligible.
	Finally, that in the limit of highly-energetic fast ions, their response approaches that of the ``dilution'' model; in particular, alpha particles are expected to have little, if any, stabilizing effect on plasma turbulence.
	We support these conclusions through detailed linear and nonlinear gyrokinetic simulations.
\end{abstract}

\maketitle
%\ioptwocol

%{\color{blue} Comments for Aylwin}
%{\color{red} Comments for everyone}

\section{Fast ion stabilization of ITG microturbulence} \label{introsec}

Microturbulence fundamentally limits the confinement time in current and future tokamak experiments~\cite{progiterphysics2}. The very gradients that are required to achieve high central densities and temperatures also provide a source of free energy. This free energy drives transport of particles, momentum, and energy, usually far in excess of collisional transport. The ion temperature gradient (ITG) mode, for instance, limits the core temperature of tokamaks \cite{sagdeevITG,coppiITG,cowley_considerations_1991}. This turbulence occurs on the scale of the thermal ion Larmor radius $\rho_i$.

Gyrokinetics is the reduction of the Fokker-Planck kinetic equation that rigorously handles electromagnetic fields whose fluctuations vary on spatial scales similar to $\rho_i$, but on timescales much slower than the gyro-frequency $\Omega_i$ \cite{frieman1982nge,sugama1996tpa,abel_multiscale_2013}.  A multitude of computational tools have been developed to solve the nonlinear gyrokinetic equation~\cite{gs2ref,candy2003eulerian,gkwRef,jenko2001nonlinear}. 
These tools take as inputs the equilibrium magnetic field, density, and temperature profiles, and predict the ensuing turbulent fluctuations and the associated transport fluxes. 
By using experimentally-determined profiles and comparing the computed fluxes to experimentally-inferred fluxes (usually from power-balance calculations), and through more detailed comparisons of turbulence characteristics, the validity of the gyrokinetic approach can be verified \cite{holland_validation_2016}.
This has been widely demonstrated \cite{holland_advances_2011,howard_quantitative_2012,white_multi-channel_2013,gorler_flux-matched_2014}. 

However, due to the scarcity of computational resources, these matching exercises of necessity entail the use of simulations that neglect various parts of the complex physics of the experimental setup.
It was discovered during one of these exercises, in which experiments on the Joint European Torus~\cite{rebut1987jet} were analysed, that effects from non-thermal minority ions had to be accounted for.
Indeed, the predictions of the gyrokinetic codes in the absence of such ions suggested far greater transport than was observed~\cite{citrin_nonlinear_2013}. This effect -- the fast ion stabilization of ITG turbulence -- is the subject of this paper.

%However, reaching an accurate agreement 
%including the detailed fluxes in various transport channels and properties of the fluctuating fields 
%sometimes requires not only the adjustment of inputs parameters within experimental uncertainties, but the refinement of the physics fidelity of simulations. This could, for instance, require adding more ion species, allowing for electromagnetic effects, or accounting for electron-ion scale coupling.  In this work, a case is examined where, unless effects caused by the presence of fast ions are accounted for, one obtains physically unreasonable thermal heat fluxes that greatly exceed what one would expect from power balance.

%This effect is important because plasma heating is essential in achieving a fusion-burning plasma, and this heating is often effected by the creation or injection of fast ions. 
This effect is important because the presence of energetic ions are essential to sustain fusion relevant bulk temperatures.
Two external heating methods are used to supply the plasma with high-energy ions: neutral beam injection~\cite{stix1972nbi} (NBI) and ion cyclotron resonance heating (ICRH)~\cite{swanson1985icrh}. As well as the fast ions resulting from external heating, another class of fast ions, energetic alpha particles, are generated from the fusion reaction itself. In the future, it is anticipated that alpha particles will provide the majority of the heating, and thus, that the plasma will be energetically self-sustaining (i.e., a burning plasma). In the local gyrokinetic simulations used herein, the practical differences between these several classes of fast ions (as well as, of course, their mass and charge) are in their distribution in phase space. 
%In particular, NBI ions and alpha particles are observed to have large spatial gradients of density, whereas ICRH-heated ions are characterized by a large gradient of energy density. These differences reflect the differences in the sources of these ions. By acting as particle and heat sources, these fast ions invariably affect the plasma in several important ways which will be highlighted throughout this work. 

%In particular, it has been observed that a significant local reduction in the turbulent transport (i.e., an ``internal transport barrier'') can occur when the fast ion concentration is 
As the main medium of heat injection, fast ions are a critical component of the fusion plasma and thus worthy of study in their own right. However, above and beyond this, it has been observed that a large fast ion population can improve the plasma confinement \cite{romanelli_fast_2010}, leading to elevated core temperatures and/or densities. It has been shown in gyrokinetic simulations \cite{citrin_nonlinear_2013} that this effect is non-trivial. 
It is important to understand how and why this stabilization occurs, particularly when extrapolating to future net-power-producing fusion reactors, where alpha particles provide most of the heating. Although, like externally driven fast ions, alpha particles will contribute significantly to the plasma pressure in reactors, their density is very small ($n_\alpha \lesssim 0.01 n_e$) compared to NBI and ICRH ions. It is therefore not obvious that they will affect turbulence in a similar way.
%Although alpha particles will also contribute significantly to the plasma pressure in reactors, their density is very small $n_\alpha \lesssim 0.01 n_e$, compensated by their high energy. If alpha particles cause a significant stabilization of microturbulence, this could revolutionize the quest for fusion energy.

Owing to the potentially profound implications of this and other effects of fast ions on microturbulence, there has been widespread study of the topic. Follow-up works \cite{citrin_ion_2014,garcia_key_2015} further characterize the stabilization and examine cases both where it is weak and where it is strong. Other research has sought to determine which properties of the fast ions contribute most strongly to their effects; these properties are typically categorized as dilution (of the main ions), non-Maxwellian velocity space distributions, and electromagnetic effects. Refs.~\cite{tardini_thermal_2007,holland_progress_2012} studied the effect that dilution has on suppressing ITG turbulence, with Ref. \cite{wilkie_global_2017} presenting a reduced model for this effect (expanded further in this work). The active response of a hot Maxwellian impurity was investigated analytically \cite{liljestrom_low_1990-1} and in simulations \cite{estrada-mila_turbulent_2006}. The effects of non-Maxwellian fast ions on turbulence were initially studied using isotropic fast ion velocity-space distributions in Ref. \cite{wilkie_validating_2015} and was further generalized for anisotropic fast ions in Ref. \cite{siena_non-maxwellian_2016}. The electromagnetic effects of fast ions consist a rich topic, primarily focused on the destabilization of Alfv\'en eigenmodes (AEs) and energetic particle modes (EPMs) \cite{fu_excitation_1989,zonca_kinetic_1996}. The theory of how electromagnetic fluctuations interact with ITG turbulence is continuing to be developed \cite{kim_electromagnetic_1993,snyder_electromagnetic_2001,bass_gyrokinetic_2010,pueschel_secondary_2013,zocco_magnetic_2015,duarte_prediction_2017}. In the present work, an analytic model for the effect of fast ions on ITG turbulence (in contrast to modes which are driven by fast ions themselves) is presented.

%It has been observed in experiment and simulation~\cite{romanelli_fast_2010,citrin_nonlinear_2013} that the presence of fast ions can have a profound effect on the turbulence. These fast ions are utilized to heat the plasma and drive current in fusion experiments, and are generated either by neutral beam injection (NBI), ion cyclotron resonance heating (ICRF), or from the fusion reaction itself (alpha particles in a deuterium-tritium mixture). While these fast ions contribute significantly to the plasma pressure, even those from auxilliary heating typically have relatively low density $n_f \lesssim 0.2 n_e$, where $n_e$ is the electron density. Yet in some cases, if they are not included in gyrokinetic simulations, unphysically large heat fluxes are observed. In particular, simulations of the ITG dominated discharge 73224 of the Joint European Torus (JET) saw an order of magnitude difference in the local bulk ion heat flux between when fast ions are included and when they are not \cite{citrin_nonlinear_2013}. Only when the energetic species were included could an experimentally-relevant power balance be obtained. This phenomenon is known as the \emph{fast ion stabilization of ITG}.

We seek to explain the dominant mechanism for fast ion-induced stabilization of ITG turbulence from first principles, focusing on a case where the effect was very strong: discharge 73224 of the Joint European Torus (JET) \cite{mantica_experimental_2009}. The results of a comprehensive linear study are presented in section \ref{linearsec}, in which it is examined to what extent the stabilization can be characterized as a change in the linear growth rate. Much about the phenomenon will be learned by doing this because it allows a wider variety of high-resolution simulations to be performed that would be unfeasible in nonlinear turbulence simulations. Then, in section \ref{modelsec}, the basic problem is simplified further by approximating the fast ion distribution function in the energetic limit, which yields an approximate analytic solution to the gyrokinetic equation. With the most important elements distilled, this simplified model is inserted into Maxwell's equations, in which an effective parameter model, applicable to linear and nonlinear simulations alike, makes itself evident. The electrostatic and electromagnetic effects are modelled respectively as effective modifications to the temperature ratio $\tau = T_i / T_e$ and $\beta_e = 8 \pi n_e T_e/B^2$, parameters to which microturbulence is known to be sensitive. This model is then benchmarked against nonlinear gyrokinetic simulations and elaborated upon further.

\section{Characterizing the effect of fast ions on the linear ITG mode} \label{linearsec}

We begin by considering the simplified linear gyrokinetic system and ask whether, and how, the presence of fast ions affects the growth of unstable ITG modes that give rise to microturbulence. We accomplish this by analyzing a case in which their effect is particularly strong: JET discharge 73224 \cite{citrin_nonlinear_2013}. We will find that the presence of fast ions has a nontrivial effect on the ITG mode beyond their global effects on the plasma and their dilution of turbulence driven by the thermal ions\footnote{The ions which are approximately Maxwellian close to the electron temperature which make up most of the positive charge and drive the ITG mode will alternatively be referred to as ``bulk'', ``thermal'', or ``main'' ions throughout this work} . 
%The work presented in this section is substantially based upon the study conducted in Ref.~\cite{iantchenko_role_2017}; more exhaustive parameter scans and details can be found therein.
In this section many of the results and analyses presented are a summary of the more extended linear analysis expounded in Ref.~\cite{iantchenko_role_2017}.

When the fluctuations are small enough to treat linearly, ITG may cause the electromagnetic fields to grow exponentially. Thus we find, for example, that the fluctuating electric potential $\phi \propto e^{- i \tilde{\omega} t}$, where $\tilde{\omega} = \omega + i \gamma $, $\omega$ is the frequency (by convention, positive for waves propagating in the ion diamagnetic direction), and $\gamma$ (if positive) is the growth rate. In this section, we repeatedly calculate this growth rate because, although eventually the fields grow until the nonlinear interaction between modes competes with the linear physics \cite{barnes_critically_2011} and the system reaches a saturated turbulent state, the linear growth rate has a direct impact on the strength of the saturated turbulence \cite{balescu_aspects_2005}. Indeed, it is often used for the estimation of saturated field amplitudes in reduced transport models \cite{staebler_theory-based_2007,bourdelle_core_2016}. 

Now in order to study the ``effect of fast ions on the linear growth rate'', it is necessary to somewhat artificially isolate particular effects because, without generating fast ions, the discharge would be fundamentally different. In other words, since fast ions are used to heat the plasma, drive the current, and act as a particle source, it is not immediately clear how to compare cases ``with'' and ``without'' fast ions. In this section, we define this comparison as being between a baseline case that includes fast ions and a hypothetical equivalent case in which fast ions do not participate in the turbulence, but in which global properties of the main ions and electrons are nevertheless the same, except for constraints imposed locally by the absence of fast ion charge.

To be effective at heating, the fast ion pressure ought to be comparable to the thermal pressure. Therefore an additional effect on the magnetic geometry is expected. Our observations indicate that fast ions have little effect on the flux surface shape, but have a non-trivial impact on the safety factor, magnetic shear, and Shafranov shift, which are all known to play a significant role in microturbulence. For more details on the stabilizing role of the fast ion pressure gradient, the reader is directed to Refs. \cite{romanelli_fast_2010,garcia_key_2015,iantchenko_role_2017}. Henceforth, in order to to isolate the effect of fast ions on the local ITG mode, the equilibrium plasma parameters, including the safety factor, magnetic shear, and flux surface shape, will remain fixed.

\subsection{Gyrokinetic framework and baseline case}

Apart from these global effects, there remains a nontrivial effect of fast ions on the ITG mode as manifest in local gyrokinetics. The simulations presented in this work were performed with the \textsc{gs2} code~\cite{kotschenreuther_comparison_1995,dorland_electron_2000}, which solves the gyrokinetic equation~\cite{frieman_nonlinear_1982}:
\begin{eqnarray} \label{gkeqn}
   \pd{h_s}{t} +& v_\parallel \mathbf{b} \cdot\nabla h_s + \vd \cdot \nabla h_s + \mathbf{v}_\chi \cdot \nabla h_s  - C\left[ h_s \right]  \nonumber\\
      &= - Z_s e \pd{\gyav{\chi}}{t} \pd{F_{0s}}{\mathcal{E}} - \mathbf{v}_\chi \cdot \nabla F_{0s}
\end{eqnarray}
for the perturbed distribution function for several \corr{isotropic} species $s$: $\delta f_s = Z_s e \phi \pd{F_{0s}}{\mathcal{E}} + h_s$. \corr{While we will focus on simulations of isotropic fast ions in this work, we will occasionally discuss the implications of anisotropy. In this case, the gyrokinetic equation is the same, but there is an additional contribution to $\delta f_s$ from the possible $\mu$-dependence of $F_{0s}$.}
The mass and charge of the species are $m_s$ and $Z_s e$ respectively. The fluctuating fields $\phi$ and $\mathbf{A}$ (the vector magnetic potential) are represented by a scalar electromagnetic potential $\chi \equiv \phi - \mathbf{v} \cdot \mathbf{A}/c $, the equilibrium distribution of species $s$ is  \corr{$F_{0s}\left(\mathcal{E},\mu, \sigma_\parallel\right)$}, and $h_s$ is the non-adiabatic part of the fluctuating distribution function which does not depend on gyrophase $\vartheta$. The equilibrium magnetic field has magnitude $B$ and points in the direction of the unit vector $\mathbf{b} \equiv \mathbf{B}/B$. The parallel velocity is defined as $v_\parallel = \sigma_\parallel \sqrt{2 \left[\mathcal{E} - \mu B\left(\theta \right) \right]/m_s}$, with energy $\mathcal{E}$, \corr{the exactly-conserved magnetic moment $\mu$}, and sign $\sigma_\parallel = \pm 1$. For a Maxwellian species with a temperature $T_s$, the thermal speed $v_{ts} \equiv \sqrt{2 T_s / m_s}$, and for a non-Maxwellian species, this represents a characteristic speed defined using an effective temperature $T_f^* \equiv - n_f \left[ \int \left(\partial F_{0f} / \partial \mathcal{E} \right) \,\dv  \right]^{-1}$. The characteristic Larmor radius is given by $\rho_s \equiv v_{ts}/\Omega_s$, where $\Omega_s \equiv Z_s e B /m_sc$. The tokamak minor radius $a$ provides an approximate length scale on which the equilibrium and fluctuations parallel to the magnetic field vary, and $\rho_* \equiv \rho_i/a \ll 1$. The quantity $\vd$ is the magnetic drift velocity of the guiding center and $\vchi \equiv \left(c/B \right) \mathbf{b} \times \nabla \gyav{\chi}$ includes the $\mathbf{E} \times \mathbf{B}$ drift along with the drifts and streaming associated with the fluctuating magnetic field. The gyro-average at fixed guiding center $\mathbf{R}$ is denoted by $\gyav{\phi} = \int_0^{2 \pi} \phi\left( \mathbf{r} \right)\, \mathrm{d}\vartheta / 2 \pi$. For simplicity, gradients of equilibrium plasma flows are ignored and the equation is solved in the frame rotating with the plasma. 
   \
A conservative linearized Fokker-Planck collision operator $C$ \cite{abel_linearized_2008,barnes_linearized_2009} is employed to model collisions of ions and electrons, and of each species with themselves (collisions between ions of different species are omitted). 

Throughout this work, we will focus on JET discharge 73224 around the flux surface with a half-width of $r=0.375 a$ (where $a$ is the half-width of the last closed flux surface). The nominal parameters for this baseline case are based on those from Ref.~\cite{bravenec_benchmarking_2016} and are listed in Tables~\ref{parametertable} and \ref{lincasetable}.

The resolution for the linear simulations of this section are as follows: there are \corr{58 grid points along the field line per poloidal turn} (parametrized by poloidal angle $\theta$), and the parallel domain extends to $\theta = \pm 22 \pi$. The velocity space resolution is 36 grid points in energy and 46 in pitch angle. The time step is 0.03 $a/v_{ti}$, run until the complex frequency is converged to within a factor of $2\times 10^{-4}$. 
There are two species of fast ions present in these simulations, each represented as a high-temperature Maxwellian. The local pressure gradient, used to rescale the local geometrical parameters according to the Miller prescription, is calculated consistently depending on the local fast ion pressure gradient.
%Given a global profile of the plasma properties, this produces a flux of particles, momentum, and energy which then feed back and adjust the plasma profile on a much longer transport timescale \cite{abel_multiscale_2013}. To ``stabilize'' the turbulence means to allow stronger gradients given a plasma heating profile, or conversely, to result in suppressed fluxes given a fixed radial plasma profile. 

\begin{figure}
   \begin{center} \includegraphics[width=0.3\textwidth]{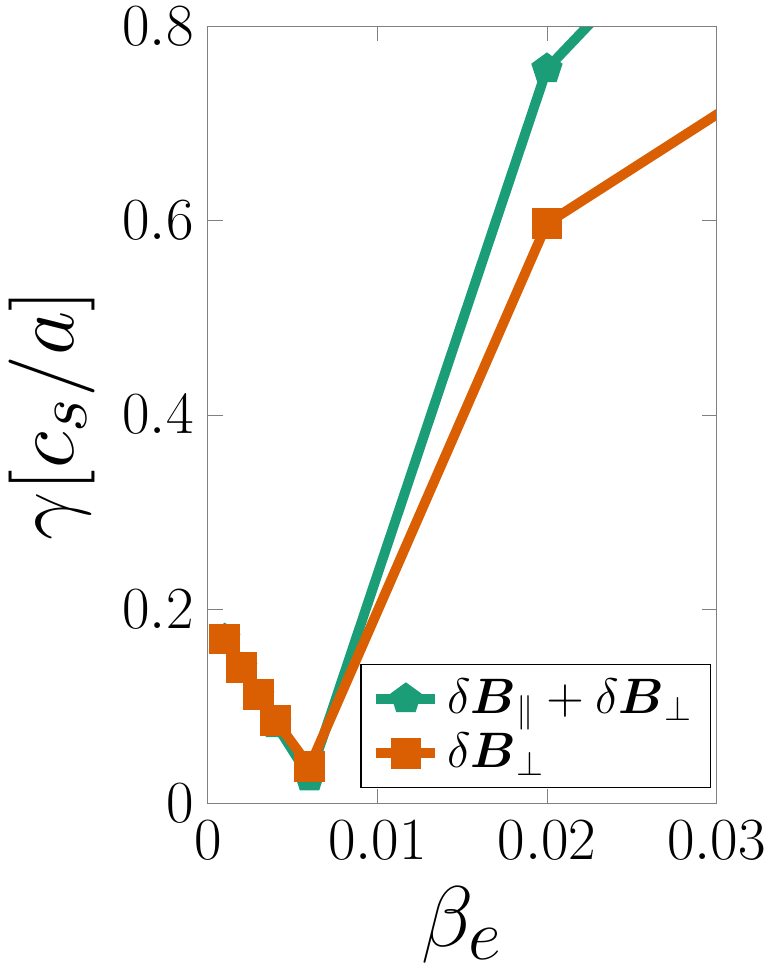} \end{center}
   \caption{\label{dbpar} Comparing ITG/KBM growth rates with fast ions with (green pentagons) and without (orange squares) compressive magnetic fluctuations (case B0 of Table~\ref{lincasetable}) at $k_y \rho_i = 0.57$.}
\end{figure}

Unless otherwise stated, all simulations include perpendicular magnetic fluctuations ($A_\parallel$), with compressive fluctuations ($\delta B_\parallel$) artificially disabled. This is justified by the relatively low $\beta_e$, and by the growth rates shown in Fig.~\ref{dbpar}. At a critical $\beta_e$, the kinetic ballooning mode (KBM) becomes unstable and the growth rate significantly increases. Only then is there a discernible difference when including $\delta B_\parallel$. As long as the mode is ITG-like, increasing $\beta_e$ is stabilizing and only the fluctuations of $A_\parallel$ need to be considered. For a more detailed treatment on the effect of compressive fluctuations on ITG modes, see Ref. \cite{zocco_magnetic_2015}.

\subsection{Fast ion-induced stabilization of the ITG mode: general observations}

\begin{figure}
   \begin{center} \includegraphics[width=0.5\textwidth]{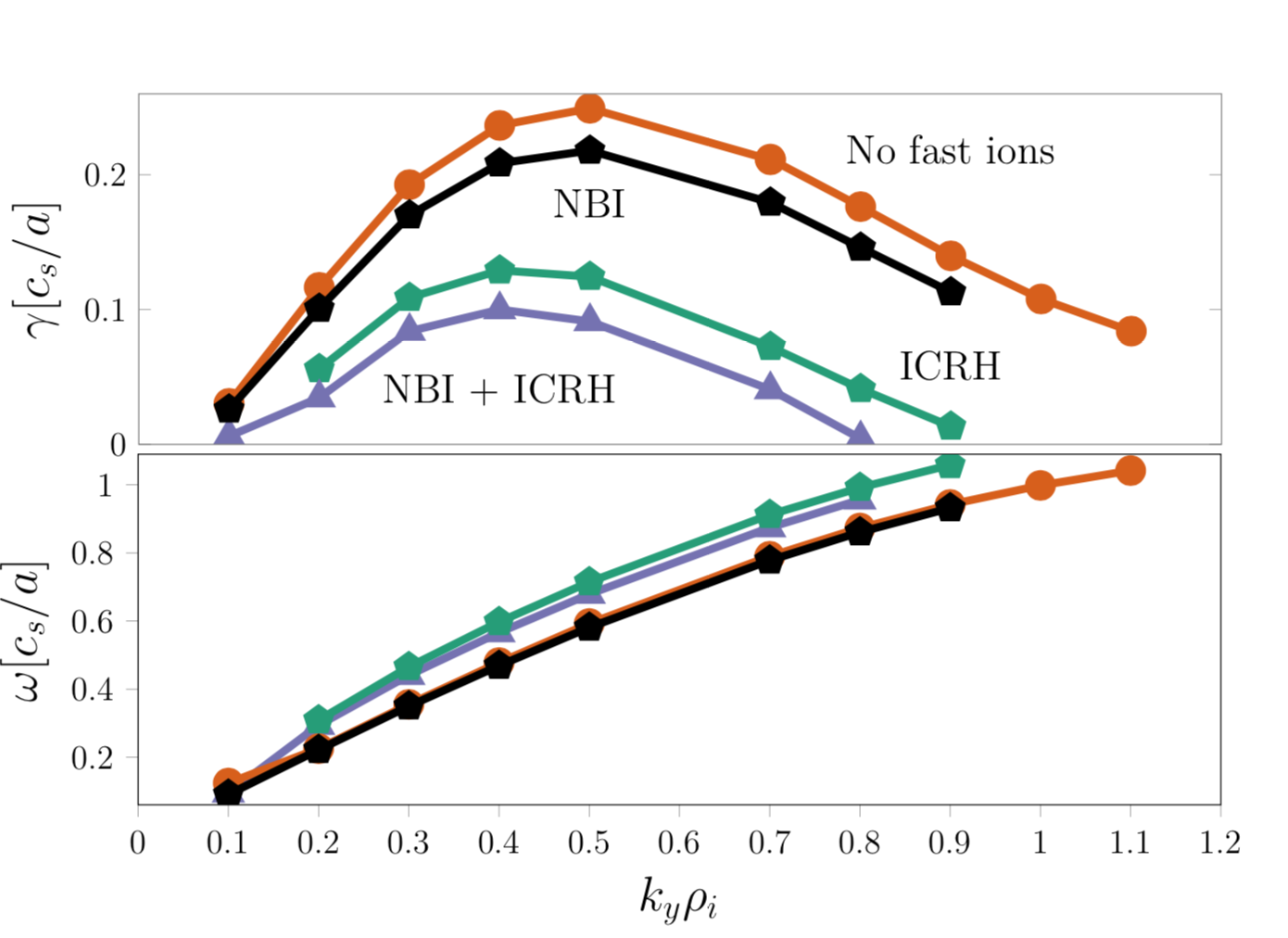} \end{center}
   \caption{\label{gammaspectrum} ITG growth rate and frequency spectra with different concentrations and types of fast ions. Orange circles show the case without any fast ions (case BXe of Table~\ref{lincasetable}), black/green pentagons show a case with only the NBI/ICRH fast ions with $Z_f n_f = 0.06/0.14$ (cases BN/BI, respectively), and violet triangles show the baseline case (B0) with both NBI ($Z_f n_f = 0.06$) and ICRH ($Z_f n_f = 0.14$). }
\end{figure}

The effect of fast ions on the ITG mode growth rate and frequency in this case is shown in Fig.~\ref{gammaspectrum}. The baseline case has both NBI fast deuterium and ICRH fast helium-3, with both types together consisting 20\% of the positive charge. Then, the growth rates were recalculated for a case with the different types of fast ions removed. A significant increase in the ITG mode growth rate is observed, with a modest change in the frequency spectrum. Finally, the both fast ions were also removed to generate a case ``without fast ions''. This shows that the effect of NBI is relatively small compared to ICRH. While the growth rate is sensitive to the presence of ICRH fast ions, the ITG mode frequency and eigenfunction are not significantly altered by the presence of fast ions, as was also observed in Ref. \cite{bravenec_benchmarking_2016}. 

A ``mixing-length'' estimate for the bulk ion heat flux based on the linear physics is $q_i \sim \gamma/k_\perp^2$. Using this estimate, our results suggest about a factor of 2 increase in the turbulence amplitude when fast ions are removed. This alone does not account for the strong nonlinear effect in Ref.~\cite{citrin_nonlinear_2013} and later in Fig.~\ref{stabexample}. These latter results showed about a factor of 10 increase in the thermal ion heat flux. It is in this sense that we say the effect of fast ions on turbulence cannot be explained in the context of linear gyrokinetics.

\iffalse
\begin{figure}
   \includegraphics[width=0.4\textwidth]{eigenfunctionses.pdf} 
   \includegraphics[width=0.4\textwidth]{eigenfunctionsem.pdf} 
   \caption{\label{eigenfunctions} Comparing eigenfunctions for a linear ITG mode with and without fast ions. }
\end{figure}
\fi

The decrease in the ITG mode growth rate when fast ions are included as a kinetic species, and the subsequent (disproportionate) decrease in the turbulence amplitude, is the subject of the remainder of this paper. We begin with the simplest explanation for this phenomenon: that only the \emph{indirect effects} of fast ions are responsible for the stabilization. Namely, we examine cases where the only effects of fast ions considered are their effects on the bulk ion and electron densities.

\subsection{Dilution}

On the temporal and spatial scales of interest, the plasma remains quasineutral:
\begin{equation} \label{equilqneutrality}
   \sum\limits_s Z_s e n_s = 0.
\end{equation}
When positively-charged impurities are included, $n_i \neq n_e$, and this means that the electromagnetic fields have proportionally less response to the bulk ions. Since the bulk thermal ions are responsible for the instability of the ITG mode, this can lead to a reduction in the vigor of the turbulence. This effect is known as dilution. For the same reason, dilution of the ions implies an \emph{enhancement} of electron-driven microinstabilities, such as the trapped-electron mode \cite{kadomtsevpogutseJETP} and the electron-temperature-gradient mode (ETG) \cite{dorland_electron_2000}. \corr{In contrast to thermal high-Z impurities, which are known to \emph{stabilize} ETG \cite{jenko_critical_2001}, the adiabatic response of singly-charged fast ions actually \emph{reduces} $Z_\mathrm{eff}^{\mathrm{(ETG)}} \equiv \sum_i Z_i^2 \left(n_i / n_e\right) \left(T_e/T_i\right)$  due to their high temperature.}

Taking the radial derivative of Eq.~(\ref{equilqneutrality}) gives:
\begin{equation} \label{equilqneutralityderiv}
   \sum\limits_s Z_s e \pd{n_s}{r} \equiv - \sum\limits_s Z_s e \frac{n_s}{L_{ns}} = 0.
\end{equation}
In a local simulation, this provides a second independent constraint that must be satisfied. This represents \emph{dilution on neighboring flux surfaces}, which influences the relationship between the electron and thermal ion density gradients. This also has a nontrivial effect on ITG turbulence and is included when we speak of ``dilution''. When results are thereby labelled, \corr{it means that fast ion fluctuations} are not included in the gyrokinetic simulations, but their equilibrium effect on the density of the bulk species and on the magnetic drift (through their contribution to $\partial \beta /\partial r$) are included. \corr{In other words, we keep the effects of $F_{0f}$ and $\nabla F_{0f}$, but not $\delta f_f$. Note that, in the high-energy limit (explored in Sec.~\ref{modelsec}), this is equivalent to the adiabatic approximation, where $\delta f_f \approx h_f \approx 0$.}
%When only one ion species is present, $Z_i n_i/n_e = Z_i L_{ne}/L_{ni} = 1$, but this is not true when an impurity such as fast ions are included. 

\begin{figure}
   \begin{center} \includegraphics[width=0.4\textwidth]{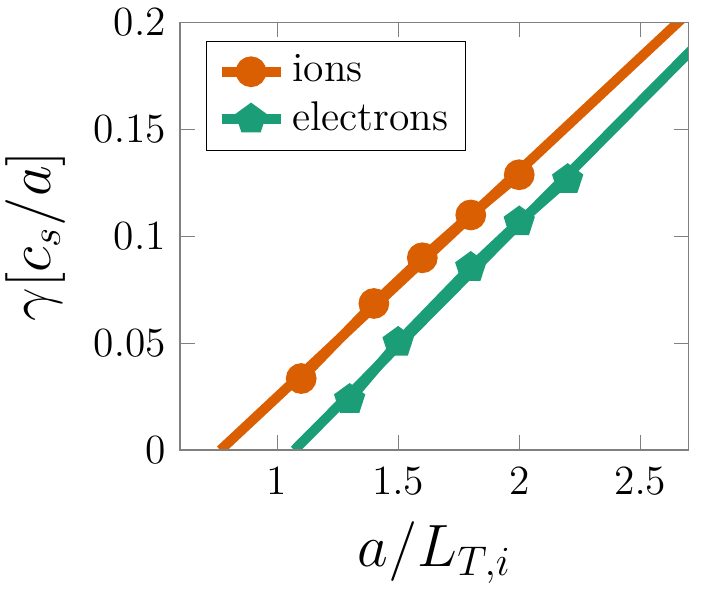} \end{center}
   \caption{\label{whichtodilute} ITG mode growth rates at $k_y \rho_i = 0.57$, as a function of the driving temperature gradient. In one case, we assume that the thermal ions replace the removed fast ions (orange circles - case BXi of table \ref{lincasetable}). In the other, electrons are removed along with fast ions (green pentagons - case BXe).}
\end{figure}

In constructing artificial cases lacking fast ions (in order to isolate and study their effect on the ITG mode and resulting turbulence), there is a choice to be made regarding whether the electron or thermal ion density (and density gradients) are changed to maintain quasineutrality. Fusion products and energetic tails heated from the bulk ion population deplete the thermal ion population. On the other hand, the physical origin of injected and minority-heated fast ions are such that they are accompanied by excess electrons.  The difference between these choices for the $k_y \rho_i = 0.57$ ITG mode is shown in Fig.~\ref{whichtodilute}. As can be seen, this choice happens to have a relatively small effect on our results, although this may not be true in general.  In this section, we choose the convention that it is the electron properties that are changed.

\iffalse
\begin{figure}
   \begin{center} \includegraphics[width=0.5\textwidth]{QLplot.pdf} \end{center}
   \caption{\label{quasilinear} Quasilinear estimate for ITG turbulent flux levels with (red) and without (red), comparing to dilution only (blue).}
\end{figure}
\fi

%By changing the bulk plasma properties and introducing additional charge of their own, fast ions also contribute to $Z_\mathrm{eff} \equiv \sum_{s\neq e} n_s Z_s^2 /n_e$, which is used in the electron collision operator. Their pressure also has an effect on the total plasma $\beta$, which affects the local magnetic equilibrium (the effect on the global magnetic equilibrium is discussed in the previous section). 

\subsection{Differentiating classes of fast ions}

\begin{figure}
   \begin{center} \includegraphics[width=0.5\textwidth]{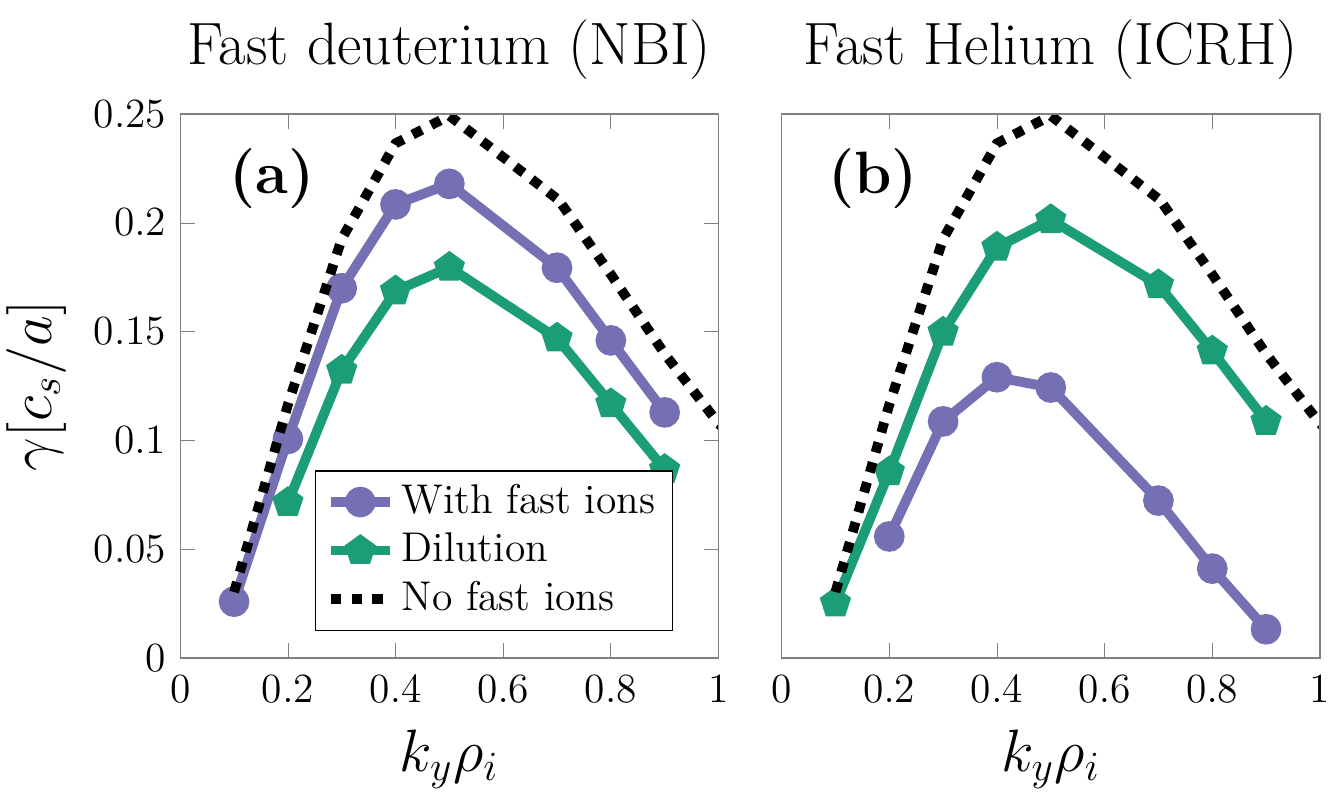} \end{center}
   \caption{\label{spectradifferentfis} ITG mode growth rate spectra when NBI and ICRH fast are included (violet circles), compared to to dilution (green pentagons) and no fast ions (black dashed). The black dashed line in both cases is case BXe of Table \ref{lincasetable}. Part (a) shows cases BN and BND, whereas part (b) shows cases BI and BID.}
\end{figure}

The different types of fast ions are characterized by whether their strong radial variation is one of particle density (``NBI-like'') or energy density (``ICRH-like''). It has previously been shown that these different classes of fast ions respond differently to turbulence when passive \cite{pusztai_turbulent_2016, wilkie_global_2017}. It is worth examining to what extent their effect on turbulence differs compared to their respective dilution effects. This contrast is shown in Figs.~\ref{spectradifferentfis}(a) and (b). The violet spectra each include one of the respective fast ion species, and the dashed black spectra are the cases without the fast ions, for which the electron density and density gradients are changed to maintain quasineutrality. Then, cases identical to the violet (``with fast ions'') were run, except that fast ions were not included as a kinetic species; only their effect on the equilibrium $n_e$ and $L_{ne}$. These ``dilution'' cases are shown in teal. We see that, relative to dilution, NBI-type fast ions are actually \emph{destabilizing}, whereas ICRH-type fast ions are more strongly stabilizing than dilution only. Therefore, the classification of fast ions, whether $L_{Tf} \gg L_{nf}$ (NBI) or $L_{nf} \gg L_{Tf}$ (ICRH), is critical to predicting and understanding their stabilizing effect. Note that alpha particles are considered NBI-like, but instead of being artificially injected, they are produced by the fusion reaction. Because the fusion source is a strong function of radius, the alpha particle density gradient is sharp.
Now the fact that the two cases with fast ions differ from their respective ``dilution only'' cases demonstrates that fast ions play a non-trivial role besides mere dilution. However, when both types are present, the (significant) kinetic effects of fast ions may cancel out and one can be left with the illusion that dilution is the dominant effect.

\begin{figure}
   \begin{center} \includegraphics[width=0.5\textwidth]{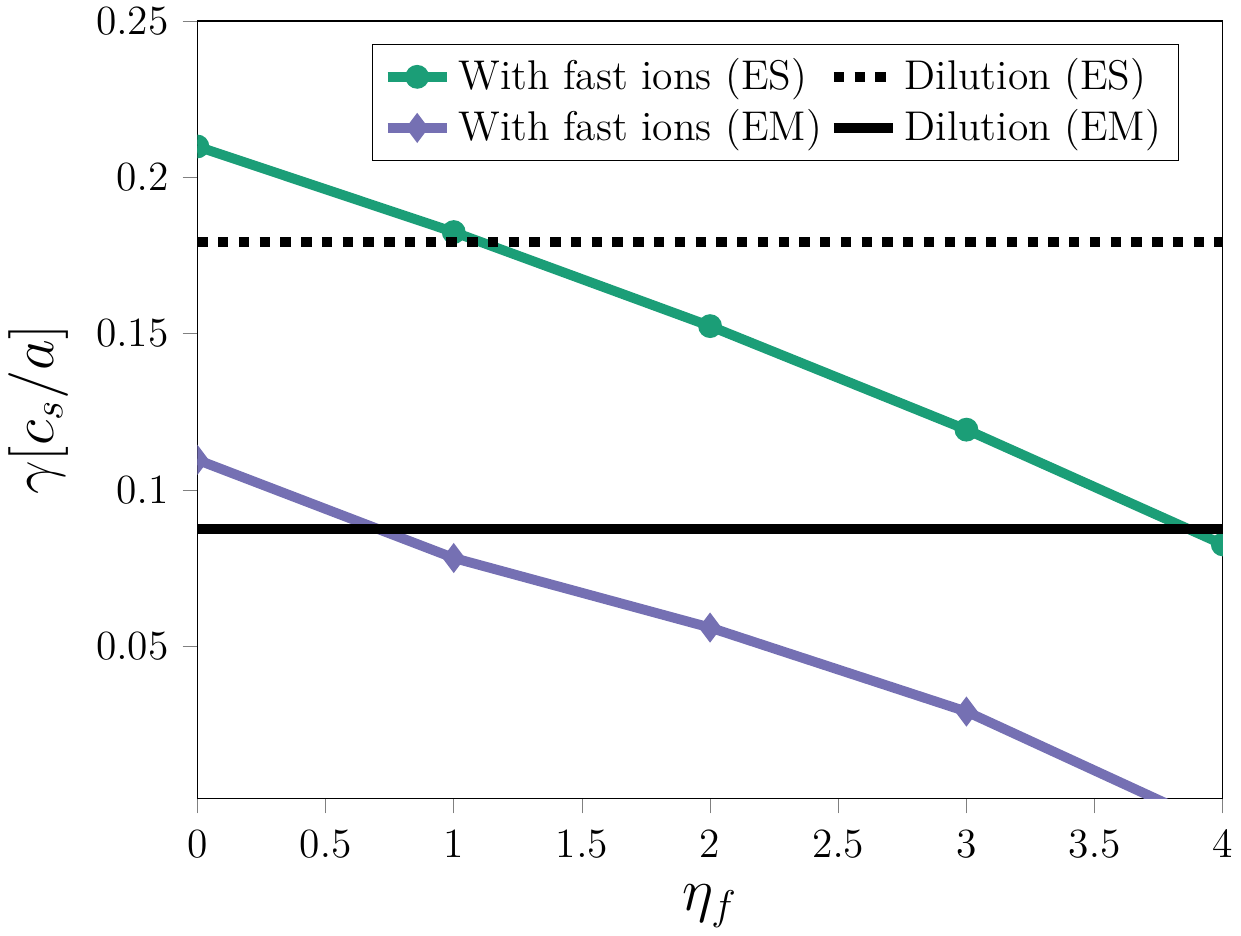}\end{center}
   \caption{\label{etascan} Scan in fast ion $\eta = T'(r)/n'(r)$ for electrostatic (teal circles) and electromagnetic (violet triangles) ITG simulations relative to electrostatic/electromagnetic dilution (dashed/solid black line, respectively) for case B0 of Table~\ref{lincasetable} at $k_y \rho_i = 0.57$. }
\end{figure}

We saw from Fig.~\ref{spectradifferentfis} that, relative to dilution, the NBI-like fast ions (strong density gradient) are destabilizing, while the ICRH-like fast ions (strong temperature gradient) are stabilizing. The parameter $\eta_f \equiv T_f'(r)/ n_f'(r) =  L_{nf}/L_{Tf}$ characterizes the relative strengths of the gradients. Since the stabilization is clearly a function of this parameter, its effect is examined in Fig.~\ref{etascan}. Here, the temperature gradient of fast deuterium is modified from the baseline case and the growth rate is compared to the case of pure dilution (black lines). Note that the dilutive effect on the bulk plasma density is not affected when changing the temperature gradient. For both electrostatic and electromagnetic cases, the threshold in the value of $\eta_{f}$ for when the fast ions become more stabilizing than simple dilution is around 0.7-1.0. This threshold will be explained with an analytic model in Sec.~\ref{taueffsec} after a model perturbed distribution function is obtained. 

\begin{figure}
   \begin{center} \includegraphics[width=0.5\textwidth]{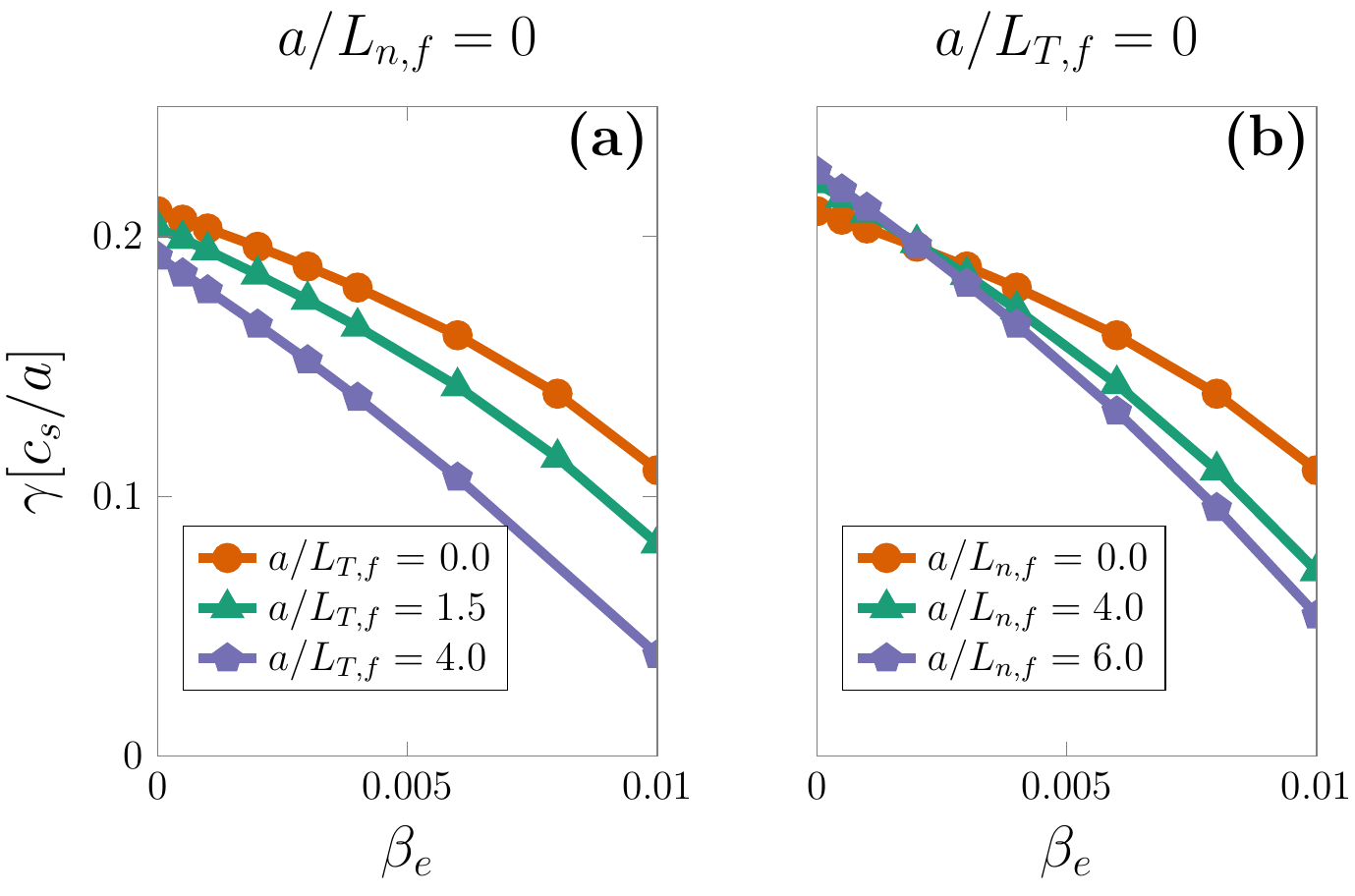} \end{center}
   \caption{\label{betascan} Scan of growth rate versus $\beta_e$ for different fast ion species: (a) ICRH-like, and (b) NBI-like, each with several different gradients of temperature and density, respectively. These are iterations of cases BI0 and BN0 of Table~\ref{lincasetable} with the fast ion gradients adjusted as indicated. For Case BN0, electron density and density gradient are likewise adjusted to maintain quasineutrality.}
\end{figure}

The effect of fast ions in local gyrokinetic turbulence simulations is often ascribed to electromagnetic effects. Here, some qualitative differences between different kinds of fast ions at different values of $\beta_e$ are catalogued. This is complicated by the fact that $\beta_e$ itself has its own electromagnetic stabilizing effect, and separating which changes to the growth rate are due to the electromagnetic stabilization and which are due to fast ions is not trivial. Therefore, scans in $\beta_e$ for the growth rate of the $k_y \rho_i = 0.57$ mode, using several different fast ion parameters, are shown in Fig.~\ref{betascan}. In all these cases, increasing $\beta_e$ is stabilizing. However, increasing the fast ion pressure gradient (either from the density gradient or the temperature gradient) does not universally decrease the ITG mode growth rate. In particular, note the case of NBI-like fast ions in Fig.~\ref{betascan}(b) ($a/L_{T,f} = 0$) at low $\beta_e$: increasing $a/L_{n,f}$ is actually net-destabilizing electrostatically. \corr{Dilution cases are not shown in Fig.~\ref{betascan}, but these still have a lower growth rate than their corresponding high-$a/L_{nf}$ cases, and is approximately equivalent in the $a/L_{nf}=0$, $a/L_{Tf}=0$ case, as expected.} Therefore, the stabilization effect of fast ions and dilution itself is somewhat sensitive to $\beta_e$ \cite{citrin_electromagnetic_2015}. The perturbed fast ion parallel current is small, but the direct effect of fast ions on the electrostatic potential could couple to $A_\parallel$, causing the behavior shown in Fig.~\ref{betascan}. 
%This is due to the corresponding change in $L_{ni}/L_{ne}$ associated with NBI-like fast ions (an effect which is weak or nonexistent in the ICRH-like case) and is considered a dilution effect. Namely, dilution on neighboring flux surfaces leads to a change in $\nabla n_e$ and/or $\nabla n_i$. It is this effect that changes depending on $\beta_e$.

\iffalse
\begin{figure}
   \begin{center} \includegraphics[width=0.5\textwidth]{critgrad_es_em.png} \end{center}
   \caption{\label{critgrad} Shift in the threshold gradient for ITG stability, comparing NBI fast Deuterium (FD) and ICRH fast Helium (He), for both (a) electrostatic and (b) electromagnetic simulations.}
\end{figure}
\fi

We have explored several potential explanations for the fast ion stabilization including: a modified linear growth rate, dilution of bulk ions, changes to magnetic geometry, and magnetic fluctuations. Although these explanations are physically motivated and in aggregate have a non-trivial effect, they are found lacking in describing the full order-of-magnitude stabilization observed in JET discharge 73224. A more fundamental treatment, valid in fully nonlinear turbulence, is thereby motivated. In the next section, a simplified model for the gyrokinetic equation, valid in the high-energy limit, will be derived and this will later be used to develop a reduced model to explain the qualitative effects presented in this section, in addition to the effect of fast ions in nonlinear simulations.

%\section{High-energy expansion} \label{expansionsec}
\section{Reduced model for fast ion effect on microturbulence} \label{modelsec}

In the previous section it was demonstrated that the effect of fast ions in stabilizing the ITG mode can go beyond dilution and depends on the details of the equilibrium fast ion phase space distribution. In this section, we examine the leading-order behavior of the gyrokinetic equation in the high-energy limit. This subsidiary expansion yields an analytic solution that applies rigorously to fast ions with a strong radial dependence. Then, this model for $h_f$ is reduced further with additional assumptions based on the ITG mode structure. When this approximate distribution function is inserted into Maxwell's equations, this leads to a physically-transparent \emph{effective parameter} model. We then discuss this model and benchmark it against linear and nonlinear gyrokinetic simulations.

\subsection{Energetic expansion of the gyrokinetic equation} \label{expansionsec}

In order to obtain a nonlinear model for the effects of fast ions, we will directly expand the gyrokinetic equation, making use of the high-energy nature of the fast ions.
Indeed, it is common to perform such expansions for electrons, taking advantage of their small mass. Commonly-used models include: drift-kinetic \cite{lee_gyrokinetic_1987,cohen:251,abel_multiscale_2013-1}, fluid \cite{lin_fluidkinetic_2001,chen_gyrokinetic_2001}, bounce-averaged \cite{gang1990nonlinear}, or adiabatic electron \cite{dorland_gyrofluid_1993} models. 
The adiabatic electron model, for example, approximates their contribution to the field equations as proportional to the electrostatic potential. In this section, an analogous model, applicable to energetic ions, will be developed. 

In the energetic limit of $\epsilon \equiv v_{ti}/v_{tf} \ll 1$, the gyrokinetic equation (\ref{gkeqn}) reads: 
\begin{equation} \label{fastgkeqn}
   v_\parallel \mathbf{b} \cdot \nabla h_f + \mathbf{v}_D \cdot \nabla h_f = - \mathbf{v}_\chi \cdot \nabla F_{0f}.
\end{equation}
In arriving at Eq.~(\ref{fastgkeqn}), ion-scale microturbulence is in mind so that the bulk ions are what set the temporal and spatial scales of the turbulent fluctuations so that $\partial h_f / \partial t \sim \omega h_f$ and $|\nabla h_f| \sim h_f/\rho_i$. The $\partial h_f / \partial t$ term, the nonlinear term, and the first term on the right hand side of Eq.~(\ref{gkeqn}) are smaller than the magnetic drift term by factors of $\epsilon^{-2}$, $\epsilon^{-3/2}$, and $\epsilon^{-3}$, respectively.  %Note that the parallel streaming term -- the first term in Eq.~(\ref{fastgkeqn}) -- is formally smaller than the magnetic drift term by one power of $\left( v_{ti}/v_{tf}\right)$, but here it is retained for generality to cover parts of phase space where $\vd$ vanishes. 
%Note that the parallel streaming term -- the first term in Eq.~(\ref{fastgkeqn}) -- is formally smaller than the magnetic drift term by one power of $\epsilon$. 
\corr{The radial gradients of the fast ion equilibrium are ordered to be strong such that $|\nabla F_{0f}| \sim \mathcal{O}\left( \epsilon^{-3/2} F_{0f}/a \right)$.
To leading order, only the magnetic drift term survives and we recover the adiabatic approximation: $h_f \approx 0$, which is equivalent to ``dilution'' in this limit. To find nontrivial effects, we therefore wish to find a solution for $h_f$ correct up to $\mathcal{O}\left( \epsilon \right)$, which is why the other terms are retained in Eq.~(\ref{fastgkeqn}) even though they are formally smaller than the magnetic drift term by one power of $\epsilon$. }

\iffalse
\begin{table}
   \caption{\label{ordertable} The ordering of each term in the gyrokinetic equation (\ref{gkeqn}) in the small parameter $\epsilon \equiv v_{ti}/v_{tf}$. }
      \begin{small}
\begin{center}
   \begin{tabular}{c c c}
      Term & Order in $\omega_D h_f$  \\
      \hline
      & \\
      $\mathbf{v}_D\cdot \nabla h_f$ & 1 \\
      $v_\| \mathbf{b}\cdot \nabla h_f$ & $\epsilon$ \\
      $\mathbf{v}_\chi \cdot \nabla F_{0f}$ &  $ \left(\frac{a}{L_{nf} + L_{Tf}} \right)\epsilon^{3/2} $ \\ 
      $\mathbf{v}_\chi \cdot \nabla h_f$ & $\epsilon^{3/2}$ \\
      $\pd{h_f}{t}$ & $\epsilon^2$ \\
      $C\left[ h_f \right]$ & $\epsilon^3$ \\
\end{tabular}
\end{center}
      \end{small} 
\end{table}
\fi

As is customary, an eikonal representation is chosen for the fluctuations \cite{antonsen_kinetic_1980} so that $\chi = \hat{\chi}\left( \theta \right) e^{iS}$, where $\mathbf{b} \cdot \nabla S =  0$ and the ballooning angle $\theta \in \left( -\infty, \infty \right)$ has been chosen as the coordinate along the magnetic field line. Hats will be dropped henceforth on $h$ and $\chi$ where there is no ambiguity. The perpendicular spatial dependence of the fluctuations is embedded in $S$, which depends on the poloidal magnetic flux $\psi$ (which labels the flux surface and is a useful ``radial'' coordinate) and a field line label $\alpha$ such that $\mathbf{B} = \nabla \psi \times \nabla \alpha$. We perform a Fourier transform in the plane spanned by $\nabla \psi$ and $\nabla \alpha$ so that the gyroaverage is represented by the Bessel function $J_{0s} \equiv J_0\left( k_\perp v_\perp / \Omega_s \right)$. Now Eq.~(\ref{fastgkeqn}) becomes:
\begin{equation} \label{altfastgkeqn}
   v_\parallel \left( \mathbf{b} \cdot \nabla \theta \right) \pd{h_f}{\theta} + i \left( \mathbf{v}_D \cdot \nabla S \right) h_f =  i c \pd{S}{\alpha} \pd{F_{0f}}{\psi} J_{0f} \chi %= -i c \pd{S}{\alpha} \pd{F_{0f}}{\psi}.
\end{equation}
This can be solved directly for $h_f$. For the integrating factor, it will be convenient to define:
\begin{equation} \label{zdef}
   z\left( \theta \right) \equiv \int^\theta_{\theta_0} \frac{\omega_D' \,\mathrm{d}\theta'}{v_\parallel' \left( \mathbf{b} \cdot \nabla \theta \right)'},
\end{equation}
where $\omega_D\left(\theta\right) \equiv \vd \cdot \nabla S$. The primed variables denote which $\theta$ is the independent variable such that $\omega_D' \equiv \omega_D\left( \theta' \right)$ (similarly for $v_\|,\mathbf{b}\cdot\nabla\theta$, $\chi$, $J_0$, and $z$). The lower limit of integration, $\theta_0$, is defined to be a point where $\omega_D$ vanishes such that $\theta_0 < \theta$ if $v_\parallel > 0$, and $\theta_0 > \theta$ if $v_\parallel < 0$.  Equation~(\ref{altfastgkeqn}) can then be rewritten as
\begin{equation}
   \pd{ }{\theta} \left( e^{i z} h_f \right) = e^{iz} i c \pd{S}{\alpha} \pd{F_{0f}}{\psi} \frac{J_{0f} \chi }{v_\parallel \mathbf{b} \cdot \nabla \theta},
\end{equation}
whose solution is:
\begin{equation} \label{hsoln}
   h_f = i c  \pd{F_{0f}}{\psi} \pd{S}{\alpha} \int_{-\sigma_\parallel \infty}^\theta J_{0f}' \chi' e^{i\left(z' - z\right)} \frac{1}{\left( \mathbf{b} \cdot \nabla \theta \right)'}\frac{\intd{\theta'}}{v_\parallel'} .
\end{equation}
%The range of integration is  $ \left(-\infty, \theta \right)$ for $v_\parallel > 0$, and $\left(\theta, \infty \right)$ for $v_\parallel < 0$ to maintain causality. 
This solution is largely inspired by that of Ref.~\cite{kim_electromagnetic_1993} and can alternatively be derived directly therefrom. The equilibrium distribution can be factored out because $F_{0f} = F_{0f}\left(\mathcal{E},\mu,\sigma_\|\right)$ is not a function of $\theta$ in these coordinates. Equation~(\ref{hsoln}) is the full solution of the gyrokinetic equation for fast ions correct to $\mathcal{O}\left( v_{ti}/v_{tf}\right)$. It is a complete linear model in the sense that it is the furthest one can take the $\left(v_{ti}/v_{tf}\right)$ expansion linearly; the formally next-largest term that would appear in Eq.~(\ref{fastgkeqn}) is the electromagnetic nonlinear term $-\left(v_\parallel/B\right) \mathbf{b} \times \nabla \gyav{A_\parallel}$.
%\begin{equation} \label{zdiff}
%   z' - z = \int^{\theta'}_\theta \frac{\omega_D(\theta') \,\mathrm{d}\theta'}{v_\parallel' \mathbf{b}' \cdot \nabla \theta'}
%\end{equation}
%is the only way $z$ appears in Eq. \ref{hsoln} (the notation $\omega_D = \vd \cdot \nabla S$ has been defined).

Without further analysis, we can read off one important consequence of this solution for $h_f$: there is no zonal reponse of the fast ions. 
This follows from the fact that the zonal modes are defined as those without any $\alpha$ variation (in the context of gyrokinetic simulations, this is often written as $k_y=0$). Hence, for such modes $\partial S\left/ \partial \alpha \right. = 0$, and so $h_f\left( k_y = 0 \right) = 0$.
Zonal flows (which arise from zonal fluctuations of $\phi$) are well known to be critical in the nonlinear saturation of ITG turbulence. 
Our approximate solution shows that fast ions have no direct impact on the zonal fields. To demonstrate the robustness of this result, simulations have been performed in which $h_f\left(k_y=0 \right) = 0$ is artificially enforced. There, no discernible difference in the nonlinear fluxes was found; see Fig.~\ref{stabexample}. 
%However, this does not rule out changes to the zonal flows \cite{citrin_electromagnetic_2015} due to nonlinear interactions with modes that fast ions \emph{do} impact.

This result is not without consequence. The saturated level of turbulence is determined by a balance between ``drive'' and zonal flows, which interact nonlinearly. \corr{Fast ions only directly affect the former and not the latter. Decreasing the drive will allow stronger zonal flows~\cite{citrin_electromagnetic_2015} and an overall damping of the turbulence: moreso than if all modes (including the zonal modes) were directly damped. 
%Suppose a saturation rule\footnote{Here, this term refers to approximations of the saturated flux levels, based at least partially on the linear growth rate spectrum, in the spirit of Ref. \cite{staebler_theory-based_2007}. } is a good approximation for a particular case that does not contain fast ions. Thus, the competition between the primary and parasitic instabilities (which generate and limit the zonal flows) is balanced in such a way that the linear growth rate is a suitable proxy for the saturated turbulent fluxes. Now, introducing fast ions will upset this balance -- we have shown in Section 2 that the fast ions will change the linear spectrum, but we have also shown that they will not change the zonal flows.
%The balance that made the saturation rule a good approximation is now broken. 
%In other words, fast ions' effect on the ITG growth rate is not balanced by a corresponding decrease in the zonal flow amplitude. 
%Therefore, if a saturation rule was a good approximation without fast ions, it would overestimate the turbulent fluxes (underestimate the fast ion stabilization) in a case with significant fast ion populations. 
A similar mechanism applies to an adiabatic electron response, which also vanishes for the zonal mode~\cite{dorland_gyrofluid_1993}, and this makes turbulence remarkably sensitive to the ion-electron temperature ratio~\cite{petty_dependence_1999}.} This is a possible explanation for the nonlinear enhancement of the fast ion stabilization.
Zonal flows play an important role in the story, but it is precisely \emph{because} fast ions do not have a zonal response.

\subsection{Simplifying the model}

The rigorous solution, Eq.~(\ref{hsoln}), still retains too much physics to be a useful reduced model. We can make further approximations which allow us to write $h_f \propto \chi$. When moments of $h_f$ are thereby taken, the proportionality factors become \emph{response functions}, which we will find very useful in interpreting the contribution of fast ions to the fluctuating electromagnetic fields.

Integrate Eq.~(\ref{hsoln}) directly by parts to obtain the approximate solution:
\begin{eqnarray} \label{hballoon}
   h_f &= c \pd{S}{\alpha} \pd{F_{0f}}{\psi} \left[ \frac{J_{0f} \chi}{\omega_D} - \int_{-\sigma_\parallel\infty}^\theta e^{i(z'-z)} \pd{ }{\theta'} \left( \frac{J_0' \chi'}{ \omega_D'} \right) \,\mathrm{d}\theta' \right]\nonumber \\
    &\approx c \pd{S}{\alpha} \pd{F_{0f}}{\psi} \frac{1}{\omega_D} J_{0f} \chi,
\end{eqnarray}
where the approximation is made because the second term is smaller by one power of $\epsilon$. A difficulty with this approximation is that $\omega_D$ vanishes at specific $\theta$ points. We thus need to assume that $\chi$ vanishes sufficiently rapidly away from $\theta=0$ that the contributions from such resonances are negligible due to the smallness of $\chi$. This is referred to as the ``strongly ballooning'' or ``outboard mid-plane localization'' approximation. It is valid to the extent that the ITG mode structure peaks at the outboard midplane, which is typically the case.  In practice, this is equivalent to approximating the integral $\int_{-\infty}^\theta \approx \int_{\theta_0 + \Delta \theta}^\theta$, stopping just short of the nearest resonance point $\theta_0$. 
Alternatively, the solution (\ref{hballoon}) could also be found by ignoring the (formally small) parallel streaming term in Eq.~(\ref{fastgkeqn}), promoting $\nabla F_{0f}$ by an additional order in $\epsilon$, and solving for $h_f$ algebraically. 

The approximation in Eq.~(\ref{hballoon}) is supported by the fact that it exhibits the correct energy dependence that we expect from the usual scalings of energetic particle transport in electrostatic and electromagnetic turbulence \cite{hauff_electrostatic_2009,zhang_scalings_2010,pueschel_anomalous_2012,wilkie_validating_2015}. 
\corr{The leading term in Eq.~\ref{hballoon}, which we will use for the fast ion response to fluctuating fields, does not contribute to the turbulent transport because it is in-phase with $\chi$. The correction from the second term results in a nontrivial phase factor, which does contribute to turbulent transport with the correct energy dependence.} 

%Now that a model distribution function is derived, we proceed to approximate it further in developing a reduced fast ion model.

\subsection{Model fast ion response function}

In this section, the model distribution function of the previous section is reduced further to remove the $\theta$ dependence, taking $\theta=0$ as the only relevant location for $\omega_D$ \corr{and $J_{0f}$}. In this case, the $v_\parallel^j$ moments of $h_f$ become related to numerical parameters $R_{jf}$, which can be transparently interpreted in the gyrokinetic field equations.

First, consider the relevant Maxwell's equations in the gyrokinetic limit: $\sum_s Z_s \delta n_s = 0$ (Poisson's equation/quasineutrality) and $\nabla_\perp^2 A_\parallel = -\left(4 \pi / c \right)\sum_s \delta j_s$ (the parallel component of Ampere's law with $\delta j_s \equiv Z_s e \int \gyavalt{\delta f_s} v_\| \dv$ being the contribution of each species to the perturbed parallel current). Assume the plasma consists of a thermal ion species, electrons, and fast ions. Considering the low mass and high thermal speed of electrons, their contribution to the perturbed current dominates over the ions. At low $\beta_e$, the electron contribution to the perturbed charge density is approximately their adiabatic response to the electrostatic potential: $\delta n_e = \left(n_e e / T_e\right) \left(\phi - \fsav{\phi} \right)$, where $\fsav{ }$ denotes the flux surface average. The field equations become: 
\begin{equation}  \label{qneutralityorig}
   \hspace{-15mm}   \phi \left( \frac{Z_i^2 e^2 n_i}{T_i} + \frac{e^2 n_e}{T_e} + \frac{Z_f^2 e^2 n_f}{T_f^*} \right) - \frac{n_e e^2}{T_e} \fsav{\phi} = Z_i \int J_{0i} h_i \,\dv + Z_f \int J_{0f} h_f \,\dv,
\end{equation} 
\begin{equation} \label{ampereslaw}
   \frac{c k_\perp^2}{4 \pi } A_\parallel = -e \int J_{0e} h_e v_\parallel \,\dv + Z_f e \int J_{0f} h_f v_\parallel \, \dv.
\end{equation}
In Eq.~(\ref{ampereslaw}), it was assumed that the thermal ion contribution to the parallel current is small compared to that of the electrons or the fast ions. \corr{Equation~(\ref{ampereslaw}) is not inconsistent with employing the adiabatic electron approximation in Eq.~(\ref{qneutralityorig}). This is because the leading adiabatic behavior of electrons, even if dominant at low $\beta_e$, does not contribute to the parallel current, while non-adiabatic corrections do contribute.} With singly-charged bulk thermal ions and a temperature ratio $\tau \equiv T_i / T_e$, the non-zonal components of quasineutrality (we have already established that the zonal fast ion contribution is negligible) can be written:
\begin{equation} \label{qneutrality}
   \phi \left( \frac{n_i}{n_e} + \tau + Z_f^2 \frac{n_f}{n_e} \frac{T_i}{T_f^*} \right) = \frac{T_i}{e n_e} \left[  \int J_{0i} h_i \,\dv + Z_f \int J_{0f} h_f \,\dv \right]
\end{equation}

To capture the gyrokinetic effect of fast ions in response to the fluctuating fields, let us define the \emph{response functions}:
\begin{equation}
   R_{jf} = Z_f \frac{c T_i}{e n_e} \pd{S}{\alpha} \int \left(\frac{v_\parallel}{v_{ti}} \right)^j \frac{1}{\omega_{D0}} J_0^2 \left( \frac{k_{\perp_0} v_\perp}{\Omega_{f0}} \right) \pd{F_{0f}}{\psi} \,\dv,
\end{equation}
where $\omega_{D0} \equiv \omega_D(\theta=0)$ \corr{(similarly, $k_{\perp0}$ and $\Omega_{f0}$ take their values at $\theta=0$}). Here, the mid-plane localization assumption of the previous section was taken further to assume that $\vd$ takes its values at the outboard midplane. This assumption is also applied to $v_\parallel$, $\mathbf{b} \cdot \nabla \theta$, and $\Omega_f$, which is equivalent to taking the large aspect ratio approximation and ignoring trapped particle effects. Up until now, the magnetic geometry and equilibrium fast ion distribution are general. However, for the sake of straightforward parametrization, let us make the further assumptions of circular geometry and Maxwellian fast ions:
\begin{equation} \label{respfcn}
   \hspace{-20mm} R_{jf} \approx Z_f^2 \frac{T_i}{T_f} \frac{n_f}{n_e} \frac{R}{2L_{nf}} \int \left(\frac{v_\parallel}{v_{ti}} \right)^j \frac{ 1 + \eta_f \left[ \left(v/v_{tf}\right)^2 - \left(3/2\right)\right]}{\left(v_\parallel^2 + v_\perp^2/2\right)/v_{tf}^2 } J_0^2 \left( \frac{k_{\perp0} v_\perp}{\Omega_{f0}} \right) \frac{e^{-v^2/v_{tf}^2}}{\pi^{3/2}v_{tf}^3}\,\dv.
\end{equation}
This is the form of the response function that will be used henceforth. Note that $R_{1f}$ vanishes by odd symmetry in $v_\parallel$ when $F_{0f}$ is isotropic. This symmetry is important for the model that follows because the latter depends on the fast ions not coupling the electrostatic and electromagnetic field equations (this coupling is precisely $R_{1f}$). Even if this symmetry is violated, one could argue that the fast ion current is small (which removes their relevance in Ampere's law) and, when $\beta_e$ is small, $v_\parallel A_\parallel /c \ll \phi$, which means that the prefactor on $R_{1f}$ is small in quasineutrality, Eq.~(\ref{qneutralityorig}).

\begin{figure}
   \begin{center}
   \includegraphics[width=0.5\textwidth]{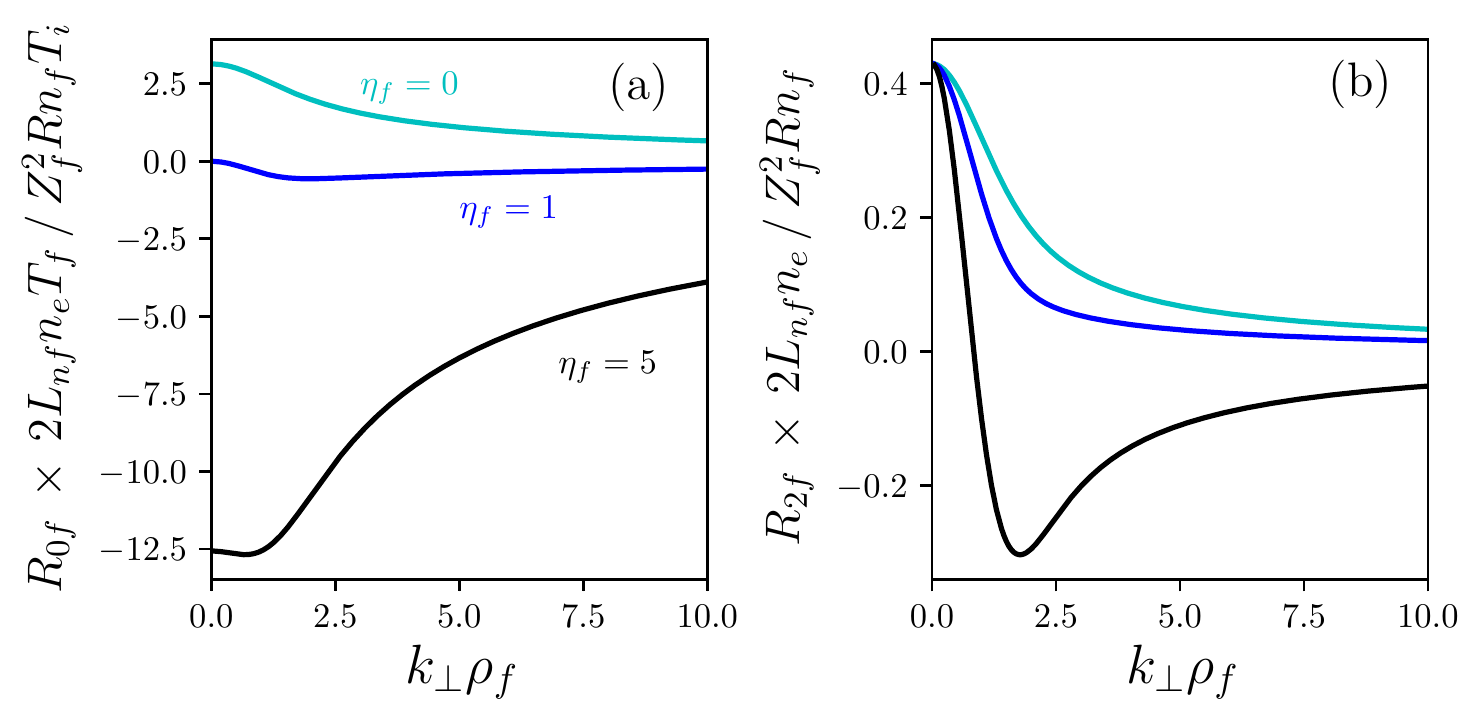}
\end{center}
\caption{\label{responsefunctions} (a) Electrostatic and (b) electromagnetic response functions for fast ions as functions of $k_\perp \rho_f$. Select values of $\eta_f = 0, 1, 5$ are shown in cyan, blue, and black respectively. The parameters that appear as prefactors in Eq.~(\ref{respfcn}) have been normalized out. Note that $R_{2f}$ is multiplied by an additional factor of $T_f/T_i$. }
\end{figure}

The response functions $R_{0f}$ and $R_{2f}$ are shown in Fig.~\ref{responsefunctions}. Note that $k_\perp \rho_f$ and $\eta_f$ consist the only nontrivial parameter dependency of the response functions; all other parameters appear as prefactors in Eq.~(\ref{respfcn}). When $|R_{0f}| \ll 1$, this means that dilution is the dominant electrostatic fast ion effect by definition. This allows us to make an important conclusion for alpha particles: unless accompanied by an unphysically strong radial gradient, the prefactor of $n_\alpha/T_\alpha$ in Eq.~(\ref{respfcn}) makes the electrostatic kinetic response of alpha particles very weak, in agreement with the results of Ref.~\cite{wilkie_validating_2015}. We will find that the electromagnetic response $R_{2f}$ has a relatively weak effect at thermal ion scales, even though it comes with an additional prefactor of $T_f/T_i$. 

\subsection{Effective parameter model} \label{taueffsec}

Having derived a simple fast ion response function in Eq.~(\ref{respfcn}), we proceed to interpret it in the context of the gyrokinetic field equations. This is done by generalizing the model presented in Ref.~\cite{wilkie_global_2017} in a way that goes beyond dilution and electrostatic turbulence. 

Consider that $h_i$ is proportional to $n_i$, but otherwise does not depend on the equilibrium ion density, except through the calculation of $\phi$ in equation~(\ref{qneutrality}). After multiplying Eq.~(\ref{qneutrality}) by $n_e / n_i$, it is seen that the same $\phi$ is ensured to be calculated from $h_i$ for all cases where the quantity $\left(n_e / n_i \right)\left[ \left(n_i / n_e \right) + \tau + Z_f^2 \left(n_f / n_e \right) \left(T_i / T_f^* \right) -  R_{0f} \right]$ is constant. Equate this quantity to a case without fast ions where $n_i = n_e$, but with an artificially defined $\taueff$. This defines an \emph{effective temperature ratio} which mimics effect of fast ions on microturbulence:
\begin{equation} \label{taueffdef}
   \taueff = \frac{n_e}{n_i} \left( \frac{T_i}{T_e} + Z_f^2 \frac{n_f}{n_e} \frac{T_i}{T_f^*} - R_{0f} \right).
\end{equation}
This is a generalization of the dilution model presented in Ref.~\cite{wilkie_global_2017}, which did not include the kinetic effect of fast ions approximated by $R_{0f}$. \corr{It is also analogous to the $\tau$ parametrization used in Ref.~\cite{jenko_critical_2001} for impurities in ETG.}
%There, nonlinear gyrokinetic simulations found excellent agreement with the known experimental scaling for thermal ion heat diffusivity from ITG: $\chi_i \sim \tau^{-3}$ \cite{petty_dependence_1999}.

A similar calculation can be performed with the parallel component of Ampere's law, Eq.~(\ref{ampereslaw}), which is rearranged thusly:
\begin{equation}
   \frac{v_{ti}}{c} \frac{e}{T_i} A_\parallel \left[ \frac{1}{2\beta_e} \frac{T_e}{T_i} k_\perp^2 \rho_i^2 + R_{2f} \right] = - e \int J_{0e} h_e v_\parallel \, \dv.
\end{equation}
By similar arguments, the same $A_\parallel$ will be calculated given an electron $h_e$ if and only if the bracketed factor is constant. This suggests an \emph{effective beta} to mimic the \emph{electromagnetic} response of fast ions:
\begin{equation} \label{betaeff}
   \betaeff = \beta_e \left( 1 + 2 \beta_e \frac{T_i}{T_e} \frac{R_{2f}}{k_\perp^2 \rho_i^2} \right)^{-1}.
\end{equation} 
The dimensionless response functions $R_{0f}$ and $R_{2f}$ are both of order unity. However, when $\beta_e$ is small, fast ions have little electromagnetic effect except the part of the spectrum where  $k_\perp \rho_i \ll 1$. Furthermore, note that dilution could also have an electromagnetic effect in Eq.~(\ref{betaeff}). This happens when the electron density, and thereby $\beta_e$, changes to maintain quasineutrality in the presence of fast ions (as was done in Sec.~\ref{linearsec}).

It is stressed that nowhere in this derivation was it assumed that the bulk plasma responds linearly to $\chi$, only that the fast ions do. This is justified from the high-energy expansion of the gyrokinetic equation. Therefore, the $\taueff$ model is expected to be valid in \emph{fully nonlinear turbulence}, and is not strictly a linear model. In cases where the fast ions are not energetic enough to take such an expansion seriously, the only recourse is to perform the fully nonlinear multi-species electromagnetic gyrokinetic simulations. The ``brisk ions'' must then be treated as a low-charge nearly-thermal impurity whose behavior is as difficult to predict as the bulk ions. But here we conclude that fast ions, to the extent that they can be classified as such, obey such an expansion and make the reduced model presented here a useful one for investigating their effect on microturbulence. We now compare this model against gyrokinetic simulations.

\subsection{Benchmarking the reduced model}
%\subsection{Qualitative predictions of the $\taueff$ model}

With the same basic plasma parameters as in Sec.~\ref{linearsec}, \textsc{gs2} simulations were performed to calculate the steady-state bulk ion heat flux. In this section, ions are used to balance quasineutrality. Therefore, in these cases, $a/L_{ne} = 0.422$, and $L_{ni}$ is determined from Eq.~(\ref{equilqneutralityderiv}). This is done to be consistent with the existing results in the literature and to avoid direct changes to $\beta_e$. The various cases presented in this section are tabulated in Appendix~\ref{paramappendix}. For nonlinear simulations simulations, the spectral range in the perpendicular direction goes up to $k_y \rho_i = 2.1$ with $k_{y,\mathrm{min}} \rho_i = 0.1$, and similarly for $k_x$. Along the field line, there are $N_\theta = 30$ grid points in each poloidal turn (of which there are 7 given the $k_x$ resolution) \corr{of the irrational flux surface}. The velocity space resolution is $N_v \times N_\lambda = 18 \times 32$. The timestep is conservatively kept below the CFL condition \cite{courant_partial_1967}, and time averages are consistently performed to be the last 60\% of the simulation: typically averaging over a period of about 300-500 $a/v_{ti}$ units. \corr{For linear simulations, we examine the $k_y \rho_i = 0.4$, $k_x = 0$ mode, and this is the mode for which we calculate $\taueff$ corresponding to the nonlinear cases.}

\begin{figure}
   \includegraphics[width=0.4\textwidth]{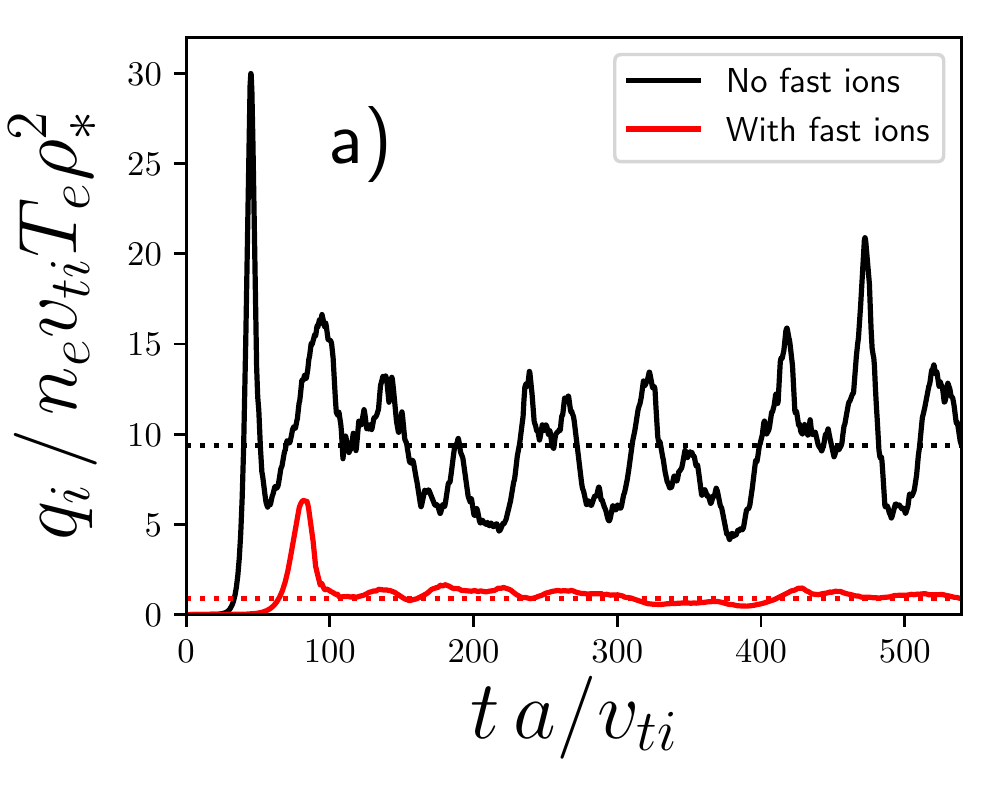}
   \includegraphics[width=0.4\textwidth]{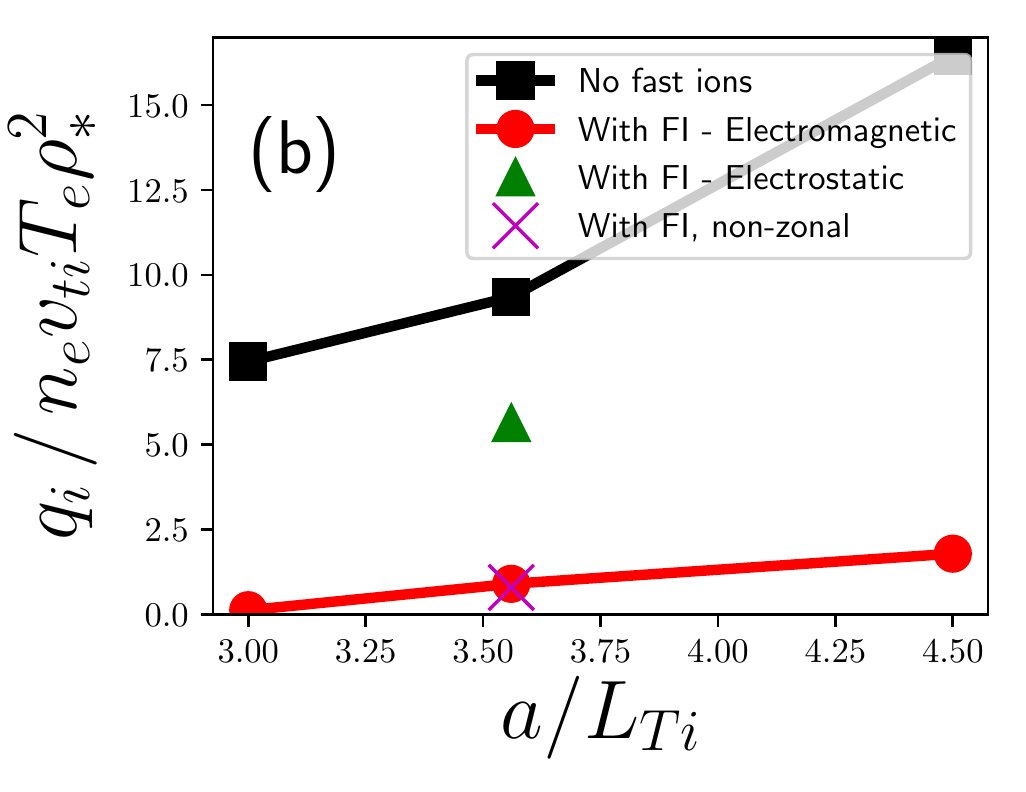}
   \caption{\label{stabexample} Examples of the fast ion stabilization of ITG turbulence in nonlinear gyrokinetic simulations of case B0 in Table \ref{lincasetable}.  (a): a time trace of thermal ion heat flux for the nominal JET 73224 discharge considered throughout this work. The time-average is shown as a dotted line. (b): the time-averaged heat fluxes for several different bulk ion temperature gradients. Green triangle is a case without $A_\parallel$ fluctuations, and the magenta $\times$ shows the case where the ``zonal'' ($k_y = 0$) component of the fast ions is artificially set to zero throughout the simulation.}
\end{figure}

Fig.~\ref{stabexample}(a) shows the time-trace of the bulk ion heat flux $q_i$ for two different simulations: the baseline case that includes fast ions and has a heat flux that approximately matches the experimental power balance, and a case with the NBI and ICRH fast ions removed. The variation of the steady-state heat flux as the thermal ion temperature gradient changes is shown in Fig.~\ref{stabexample}(b). From there, one can see that, along with a change in the critical gradient, there is also a strong reduction in the slope. Also shown is another example with fast ions, but here magnetic fluctuations were removed from the simulation. This shows that electromagnetic fluctuations are clearly stabilizing in their own right. Also shown in Fig.~\ref{stabexample}(b) is the steady-state heat flux for a case with fast ions, but with their zonal component ($h_f\left(k_y = 0 \right)$) artificially nullified. The fact that this case is nearly indistinguishable from the standard case implies that the fast ions have negligible direct effect on the zonal flows, as discussed in Sec.~\ref{expansionsec}.

We wish to use the $\taueff$ model of Eq.~(\ref{taueffdef}) to estimate the expected strength of the fast ion stabilization. To this end, we consult an empirical scaling for the bulk ion heat diffusivity $\chi_i \propto \tau^{-3}$ \cite{petty_dependence_1999}. Although it has not been derived from first principles, we can use this scaling as a useful indication for the approximate strength of fast ion stabilization based on $\taueff$. Furthermore, the presence of a carbon impurity in this case interferes with the direct calculation of $\taueff$ (see Sec.~\ref{caveatsec}). Nevertheless, for the parameters of JET discharge 73224, one obtains $\taueff = 2.9$. If the empirical scaling is to be taken seriously, this implies an even stronger stabilization than observed in Fig.~\ref{stabexample}. This indicates that the strong stabilization observed in some cases can be at least qualitatively described by the simplified model presented here. In light of this analysis, the sensitivity of microturbulence to the presence of fast ions is no mystery. In fact, the model is \emph{over}-sensitive compared to simulation.

%\subsection{Benchmarking the $\taueff$ model against nonlinear simulations}

\begin{figure}
   \includegraphics[width=1.0\textwidth]{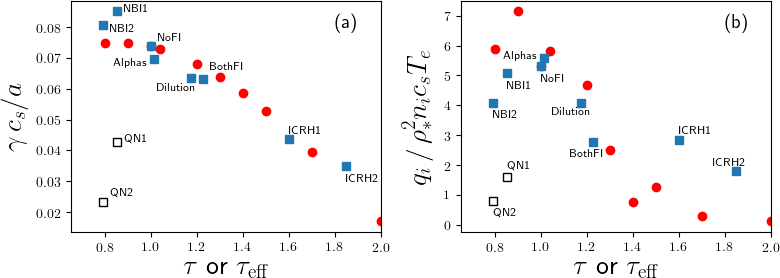}
%   \caption{\label{taueffbenchmark} Comparing the nonlinear steady-state heat flux of thermal ions when various combinations of fast ions are present (blue triangles), to the $\taueff$ model (green circles and dashed line). The latter do not include fast ions, but have $T_i$ changed to match the calculated $\taueff$. Black arrows connect to cases where $L_{ni} = L_{ne}$, which is not the case for consistent NBI-like simulations. }
   \caption{\label{taueffbenchmark} Comparing the (a) linear growth rates and (b) nonlinear steady-state heat flux of thermal ions when various combinations of fast ions are present (blue squares), to the $\taueff$ model (red circles). The latter do not include fast ions, but have $T_i$ changed to match the calculated $\taueff$. Hollow boxes (QN1 and QN2) are cases with NBI fast ions with the equilibrium ion density gradient adjusted to maintain quasineutrality. Labels refer to the parameters in Table~\ref{nlcasetable}. }
\end{figure}

The $\taueff$ model is further benchmarked with a collection of additional linear and nonlinear simulations. These cases have the same baseline parameters as those before (tabulated in Table~\ref{parametertable}), but with no carbon impurity and simplified fast ion parameters (Table~\ref{nlcasetable}). The red circles in Fig.~\ref{taueffbenchmark}(a) represent the ITG mode growth rate at $k_y \rho_i = 0.4$ with kinetic electrons and an adjusted ion temperature, but no fast ions are present, even via dilution. The simulations with fast ions include either ``ICRH-like'' ($a/L_{nf} = 0$, $a/L_{Tf} = 5$), or ``NBI-like'' ($a/L_{nf} = 5$, $a/L_{Tf} = 0$), each with a nominal density of $n_f = 0.15 n_e$ and $T_f = 10 T_e$. The cases NBI2 and ICRH2 each have $n_f = 0.2 n_e$ instead. Another case is ``alpha-like'', which have $a/L_{nf} = 4.5$, $a/L_{Tf} = 0.5$, $n_f = 0.0075 n_e$, $Z_f = 2$, $m_f = 2 m_i$, and $T_f = 200 T_e$. The case ``BothFI'' has two different fast ions species each at half density ($n_f = 0.075$) with the respective gradient length scales and temperatures listed above. For these cases with fast ions, $\taueff$ was calculated at the $k_y \rho_i = 0.4$ according to Eq.~(\ref{taueffdef}) and plotted accordingly on the horizontal axis. In most cases, the bulk plasma gradients and magnetic geometry were held fixed, despite changes in the fast ion parameters. The exception to this are the hollow boxes, in which the ion density gradient was changed consistently with the presence of NBI-like fast ions. Figure~\ref{taueffbenchmark}(b) shows the steady-state ion heat flux from the corresponding nonlinear simulations.

%As is explained below, the $\taueff$ model does not account for changes in the bulk plasma density gradient as a result of a significant fast ion density gradient. Therefore, we expect the model to be more valid for ICRH fast ions or alphas particles, both of which have low density gradients (the latter by virtue of its low density overall, even if $a/L_{n\alpha}$ is relatively large). In Fig. \ref{taueffbenchmark}, the black arrows associated with NBI cases point to cases whose $L_{ni}$ for the bulk ions was adjusted to match the other cases. This is done to indicate that this is the cause of the disagreement and is primarily responsible for the stabilization effect of NBI-like fast ions. Unfortunately, this effect is not contained in the $\taueff$ model, even in principle..

The fidelity of the $\taueff$ model is indicated by the proximity of the blue squares to the red circles in Fig.~\ref{taueffbenchmark}. The growth rates show excellent agreement, while the nonlinear heat flux captures the general trend. For the nonlinear ICRH-like cases, the $\taueff$ model significantly \corr{over-predicts the impact of fast ions}. One explanation is that these cases are on the verge of being dominated by a fast ion-driven mode, possibly a fast ion-driven ITG mode. This is made clear when the fast ion temperature is increased to $T_f = 15 T_i$ and $20 T_i$. For these modified cases, agreement with the $\taueff$ model is much better, but is not shown in Fig.~\ref{taueffbenchmark} because it is not longer thermal-ITG and the agreement is likely coincidental. \corr{Another possible contribution to the $\taueff$ model's under-prediction of the heat flux is that it makes use of the adiabatic electron model in the electrostatic field equation, while the simulations in Fig.~\ref{taueffbenchmark} included kinetic electrons. Furthermore, $\taueff$ depends on $k_\perp \rho_i$, so in nonlinear simulations, we must choose a single representative mode number if we wish to specify a single $\taueff$ parameter. } 

A prediction of the model that matches particularly well with nonlinear electromagnetic gyrokinetic simulations is that alpha particle fluctuations play little role in stabilizing ITG turbulence. The Maxwellian alpha-particle-like species shown in Fig.~\ref{taueffbenchmark} affect the bulk ion heat flux by only about 6\%. This can be explained in the $\taueff$ model by the large temperature and small density of these species. Note that high temperature, like those of alpha particles, is exactly where we expect the model to be most accurate. From Eq.~(\ref{respfcn}), one can see that $R_{0f} \rightarrow 0$ as $T_f/T_i \rightarrow \infty$, which means that dilution is the dominant effect of such fast ions. Even the electromagnetic response is small for this case: $\betaeff \approx 1.014 \beta_e$. As demonstrated for the electrostatic case in Ref.~\cite{wilkie_validating_2015}, even accounting for the non-Maxwellian nature of alpha particles has little effect on ITG-driven turbulence. 
We have neglected the changes on the magnetic geometry caused by the alpha particle pressure gradient. While this is known to have a significant impact on microturbulence, it is beyond the scope of the phenomenon we attempt to isolate in this work.

Surprisingly, in none of the cases shown in Fig.~\ref{taueffbenchmark} does $\betaeff$ depart by more than about 15\% from $\beta_e$. This is evident from Eq.~(\ref{betaeff}) since $R_{2f}$ if of order unity, but $k_\perp \rho_i \sim 1$ and $\beta_e \ll 1$. Since fast ions can play a key role in driving electromagnetic modes unstable, one would expect their influence on the ITG mode to be associated with the $\beta_e$ stabilization. Indeed, low $k_\perp \rho_i$ is relevant for Alfv\'en eigenmodes and energetic particle modes, and is where one might see a significant electromagnetic effect from fast ions. Therefore, according to the first-principles linear model, at ITG-relevant mode numbers, the effect of fast ions is mostly through their contribution to the electrostatic field equation. Their contribution to the electromagnetic field equation is precisely the perturbed fast ion current, which is small compared to that of electrons for the cases studied here. However, this does not rule out, for example, an indirect effect of $\tau$ or $\taueff$ on the magnetic fluctuations. Furthermore, if the electron density were increased to maintain quasineutrality (as opposed to decreasing the thermal ion density as was the convention followed in this section), this has a \emph{direct} change on $\beta_e$, which can have its own significant effect on the turbulence.

\begin{figure}
   \includegraphics[width=0.3\textwidth]{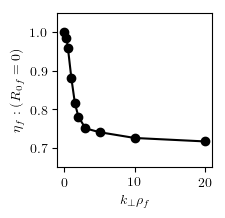}
   \caption{\label{etacrit} Threshold $\eta_f$ for which fast ions are stabilizing/destabilizing relative to dilution, as estimated from numerical calculation of Eq.~(\ref{respfcn}).}
\end{figure}

The threshold for stabilization relative to dilution is determined by the sign of $R_{0f}$. One can obtain an analytic estimate by taking the $k_\perp \rho_f \rightarrow 0$ limit in approximating the Bessel function in Eq.~(\ref{respfcn}). In this case, one obtains a threshold $\eta_f = 1$. As $k_\perp \rho_f$ becomes finite and large (as is appropriate for fast ions in thermal ion-scale turbulence), this threshold can be estimated from numerical calculation of the integral in Eq.~(\ref{respfcn}) (note that all other parameters are multiplicative factors and the threshold only depends on $k_\perp \rho_f$). See Fig.~\ref{etacrit} for these calculations. At high $k_\perp \rho_f$, the threshold approaches $\eta_f \approx 0.70$. This is in excellent agreement with Fig.~\ref{etascan}, in which the threshold was close to but less than $\eta_f =1$.

\subsection{Limitations of the reduced model} \label{caveatsec}

Many assumptions were made on the way to writing Eqs.~(\ref{taueffdef}) and (\ref{betaeff}) that it is worth considering what has been lost.

One can see from Fig.~\ref{taueffbenchmark} that the model does poorly in predicting the results of simulations with a relatively large concentration of NBI-like fast ions (the hollow black boxes). The reason for this is because they have a strong density gradient (such that $\nabla n_f \sim \nabla n_i$) and thereby have an effect on the density gradients of the bulk plasma via dilution. Note that the effect of \emph{local} dilution is included in Eq.~(\ref{taueffdef}), but that on neighboring flux surfaces is not included even in principle. To show that this is responsible for the disagreement, other simulations were run where the ion density gradient was \emph{not} changed, and these cases are also presented in Fig.~\ref{taueffbenchmark} (labelled as ``NBI'' and ``More NBI''), and there the model performs much better. Unfortunately for the model, it appears that the fast ion-induced change in $L_{ni}$ is the dominant stabilization effect of NBI-like fast ions. Although progress is being made \cite{tegnered_impact_2017}, the general theory of how turbulence scales with the density gradient, especially when that of the ions and electrons differ, is not generally known. Note that this does not jeopardize the applicability of the model to alpha particles. Even though the alpha particle gradient \emph{scale length} is short, their densities are so small that $|\nabla n_\alpha| \ll |\nabla n_i|$ and the overall effect of dilution is weak.

In order to derive Eq.~(\ref{taueffdef}), we had to ignore other impurities, such as thermal carbon or tungsten. These are known to have an impact on turbulence \cite{jenko_critical_2001,porkolab_transport_2012,pusztai_turbulent_2013,ennever_effects_2015}, but their response to ion-scale turbulent fields is not even approximately linear. Therefore, their contribution to Eq.~(\ref{qneutrality}) is difficult to predict \emph{a priori}. The simulations shown in Fig.~\ref{taueffbenchmark} did not have such an impurity, although the simulations in Fig.~\ref{stabexample} did, consistent with the experiment.

While it is undisputed that electromagnetic fluctuations alone can have a strong effect on ITG turbulence, the reduced model presented here indicates that the fast ions play little role in this phenomenon (due to the small contribution of the fast ion current relative to that of electrons). Nevertheless, fast ions are known to destabilize Alfv\'en eigenmodes \cite{fu_excitation_1989} and geodesic acoustic modes \cite{fu_energetic-particle-induced_2008}, and it is conceivable that this could play a role in interacting with microturbulence \cite{bass_gyrokinetic_2010}. Ref. \cite{garcia_key_2015} studied such a case where beta-induced Alfv\'en eigenmodes co-existed with, and had a nontrivial effect on, ITG turbulence. The $\taueff$ and $\betaeff$ models are not expected to capture the effects of modes that are driven unstable by fast ions, only their approximate effect on thermal ion-driven modes.

\section*{Conclusion}

This work presented a reduced model, derived analytically from first principles, for the effect of fast ions on ITG and other forms of ion-scale turbulence. This model is based on the high-energy limit of the gyrokinetic equation and on the ballooning structure of the ITG mode, and provides physical insight into the sensitivity of microturbulence to the presence of fast ions. Several important linear and nonlinear results were presented, which highlighted the differences between different classes of fast ions as they affect ITG turbulence. 

It was found that fast ions with strong density gradients are less stabilizing than those with relatively large temperature gradients, all else being equal. The chief stabilizing effect of the former is dilution of the thermal ions, an effect which is tempered by the \emph{destabilizing} kinetic response of these ``NBI-like'' ions. Fusion-produced alpha particles in burning plasmas fall into this category, and their effect was found to be small, owing to their low density and high energy. The strong stabilization observed with ``ICRH-like'' fast ions can be explained in light of the sensitivity of microturbulence to the thermal ion-electron temperature ratio, which acts as a proxy for the contribution of fast ions to Maxwell's equations even in cases when this contribution is not trivial.  

To reliably predict the effect of fast ions when the separation of energy is not as extreme as it is for alpha particles, multi-species nonlinear gyrokinetic simulations are required. Nevertheless, the theory presented here provides a useful estimate for the baseline effect, to which more sophisticated physics can later be added. Ideas for expanding this model include predicting the impact of: thermal impurities, changes to the equilibrium thermal plasma density gradients, and nontrivial interaction with energetic particle-driven modes.

The authors would like to thank T. F\"ul\"op, J. Citrin, M. J. Pueschel, and R. Bravenec for helpful discussions, data, and feedback. Simulations were run on the CINECA Marconi cluster. GW was supported  by the Vetenskapsr{\aa}det (VR) Framework grant for Strategic Energy Research (Dnr. 2014-5392), JET modelling task T17-04, and the EUROfusion Researcher Grant. IP was supported by the VR International Career Grant (Dnr. 330-2014-6313), and Marie Skladowska Curie Action Cofund (project INCA 600398). This work has been carried out within the framework of the EUROfusion Consortium and has received funding from the Euratom research and training programme 2014-2018 under grant agreement No 633053. The views and opinions expressed herein do not necessarily reflect those of the European Commission.

\appendix

\section{\corr{Simulation paramters}} \label{paramappendix}

Table \ref{parametertable} lists the local geometrical parameters used for all the simulations in this work. These consist of the local geometric parameters, the thermal temperature gradients, and the Carbon properties (when present). These values do not change among all the different iterations in this work, except when explicitly scanned upon (such as, for example, the scan in $a/L_{Ti}$ in Fig.~\ref{whichtodilute}).

\begin{table}
   \caption{\label{parametertable} List of local parameters used for JET discharge 73224, based on Ref. \cite{bravenec_benchmarking_2016}. }
      \begin{small}
\begin{center}
   \begin{tabular}{r c c}
      \hline
      Parameter & Symbol & Value \\
      \hline
      Flux surface half-width & $r$ & 0.375 $a$ \\
      Major radius  & $R$ & 3.12 $a$ \\
      Therm./mag. pressure ratio & $\beta_e$ & 0.0033 \\
      Safety factor & $q$ & 1.74 \\
      Magnetic shear & $\hat{s}\equiv \left(r/q\right) \partial q / \partial r$ & 0.523 \\
      Flux surface elongation & $\kappa$ & 1.26 \\
      Elongation gradient  & $a\kappa'$ & 0.105 \\
      Flux surface triangularity & $\delta$ & 0.03 \\
      Triangularity gradient  & $a\delta'$ & 0.0027 \\
      Radial change in Shafranov shift & $\Delta'$ & -0.14 \\
      Ion temperature length scale & $a/L_{Ti}$ & 3.56 \\
      Electron temperature length scale & $a/L_{Te}$ & 2.23 \\
      Impurity concentration & $n_C / n_i$ & 0.039 \\
      Impurity density length scale & $a/L_{nC}$ & 0.422 \\
      Impurity temperature length scale & $a/L_{TC}$ & 3.56 \\
       \hline
\end{tabular}
\end{center}
      \end{small} 
\end{table}

   Table \ref{lincasetable} shows the fast ion parameters, along with the effects they may or may not have on the bulk plasma (depending on the specific case being studied - some are intentionally left non-quasineutral for demonstration purposes). These are the cases used for the linear simulations in Sec.~\ref{linearsec}, while the nonlinear results for cases B0 and BXi are shown in Fig.~\ref{stabexample}. Case B0 is considered the ``baseline'' case of this work and, along with Table~\ref{parametertable} is based on the parameters reported in Ref.~\cite{bravenec_benchmarking_2016}.

\begin{table}
   \caption{\label{lincasetable} The plasma species parameters used for the various cases in Sec.~\ref{linearsec}.  Carbon has the same temperature and temperature gradient as the bulk ions, $n_c = 0.039 n_i$, $a/L_{nC} = 0.422$, and $T_C = T_C = T_e$. In general, ion density and density gradients were held fixed, except for case BXi.}
      \begin{tiny}
\begin{center}
   \begin{tabular}{l | c c c  c c c c | c c c c}
      \hline
      Case  & $n_i/n_e$ & $a/L_{ne}$  & $a/L_{ni}$   & $Z_{f1} n_{f1}/n_i$ & $T_{f1}/T_e$ & $a/L_{nf1}$ & $a/L_{Tf1}$ & $Z_{f2} n_{f2}/n_i$ & $T_{f2}/T_e$ & $a/L_{nf2}$ & $a/L_{Tf2}$  \\
      \hline
      B0 &  0.65 & 0.422  & 0.006 &  0.093 & 9.8 & 4.72 & 1.03 & 0.216 & 6.9 & 0.503 & 7.406\\
      BXe &  0.81 & 0.085  & 0.006 &  - & - & - & - & - & - & - & -\\
      BXi & 0.85  & 0.422  & 0.422  &  - & - & - & - & - & - & - & -\\
      BD &  0.65 & 0.422  & 0.006 &  - & - & - & - & - & - & - & -\\
      BN &  0.75 & 0.408  & 0.006 &  0.093 & 9.8 & 4.72 & 1.03 & - & - & - & -\\
      BND &  0.86 & 0.408  & 0.006 &  - & - & - & - & - & - & - & -\\
      BN0 &  0.65 & 0.428  & 0.006 &  0.093 & 9.8 & 6.0 & 0.0 & 0.216 & 6.9 & 0.0 & 0.0 \\
      BI &  0.69 & 0.147  & 0.006 & - & - & - & -&  0.216 & 6.9 & 0.503 & 7.406 \\
      BID &  0.94 & 0.147  & 0.006 & -  & - & - & - & - & - & - & -\\
      BI0 &  0.65 & 0.068  & 0.006 &  0.093 & 9.8 & 0.0 & 0.0 & 0.216 & 6.9 & 0.0 & 4.0  \\
       \hline
\end{tabular}
\end{center}
      \end{tiny} 
\end{table}

Finally, in Table~\ref{nlcasetable}, we tabulate the parameters used in the simulations shown in Fig.~\ref{taueffbenchmark}. This case is a simplified version of case B0, without the carbon impurity and simplified fast ion gradients. For most cases, the bulk plasma gradients were held fixed, except for the ``Alphas'' case, and cases QN1 and QN2, where $a/L_{ni}$ was adjusted for quasineutrality. 

\begin{table}
   \caption{\label{nlcasetable} The plasma species parameters used for the nonlinear cases in Sec.~\ref{modelsec}. There are no thermal impurities in these simulations. Electron properties were held fixed  for these cases: $T_i/T_e = 1$ and $a/L_{ne} = 0.422$. All fast ions are singly-charged deuterium except for the ``Alphas'' case. Labels correspond to those in Fig.~\ref{taueffbenchmark}.}
      \begin{tiny}
\begin{center}
   \begin{tabular}{ l | c c  | c c c c | c c c c}
      \hline
       Label &$n_i/n_e$ & $a/L_{ni}$   &  $ n_{f1}/n_e$ & $T_{f1}/T_e$ & $a/L_{nf1}$ & $a/L_{Tf1}$ & $n_{f2}/n_e$ & $T_{f2}/T_e$ & $a/L_{nf2}$ & $a/L_{Tf2}$  \\
       \hline
       NoFI&  1.0 & 0.422  & - & - & - & - & - & - & - & -\\
       Dilution&  0.85 & 0.422 & - & - & - & - & - & - & - & -\\
       Alphas&  0.985 & 0.360  & 0.015 & 200.0 & 4.5 & 0.5 & - & - & - & -\\
       BothFI &  0.85 & 0.422  & 0.075 & 10.0 & 5.0 & 0.0 & 0.075 & 10.0 & 0.0 & 5.0\\
       ICRH1 & 0.85 & 0.422  & - & - & - & -& 0.15 & 10.0 & 0.0 & 5.0 \\
       ICRH2 & 0.85 & 0.422 & - & - & - & - & 0.2 & 10.0 & 0.0 & 5.0 \\
       NBI & 0.85 & 0.422  &  0.15 & 10.0 & 5.0 & 0.0 & - & - & - & -\\
       NBI2 &  0.85 & 0.422  & 0.2  & 10.0 & 5.0 & 0.0 & - & - & - & -\\
       QN1 & 0.85 & -0.386  & 0.15 & 10.0 & 5.0 & 0.0 & - & - & - & -\\
       QN2 & 0.85 & -0.386  & 0.2 & 10.0 & 5.0 & 0.0 & - & - & - & -\\
      \hline
\end{tabular}
\end{center}
      \end{tiny} 
\end{table}

\iffalse
The bulk and NBI ions are deuterium, ICRH is helium-3, and the impurity is carbon-12. In Sec.~\ref{linearsec}, values of electron density and density gradient are such that quasineutrality, Eqs.~(\ref{equilqneutrality}) and (\ref{equilqneutralityderiv}), are satisfied. In Sec.~\ref{modelsec}, $a/L_{ne} = 0.422$ and the ion density and density gradients were modified.  

   \fi

\section*{References}

\bibliographystyle{unsrt}
\bibliography{zotero,ian-refs}

\end{document}